\documentclass{vldb}
\usepackage{graphicx}
\usepackage{amsmath}
\usepackage{balance}  % for  \balance command ON LAST PAGE  (only there!)

\usepackage{csquotes}

\DeclareMathOperator*{\argmax}{arg\,max}

\newcommand{\ix} {\hspace*{2em}}

\begin{document}

% ****************** TITLE ****************************************

\title{Cognitive Database: A Step towards Endowing Relational Databases with Artificial Intelligence Capabilities}

% possible, but not really needed or used for PVLDB:
%\subtitle{[Extended Abstract]
%\titlenote{A full version of this paper is available as\textit{Author's Guide to Preparing ACM SIG Proceedings Using \LaTeX$2_\epsilon$\ and BibTeX} at \texttt{www.acm.org/eaddress.htm}}}

% ****************** AUTHORS **************************************

% You need the command \numberofauthors to handle the 'placement
% and alignment' of the authors beneath the title.
%
% For aesthetic reasons, we recommend 'three authors at a time'
% i.e. three 'name/affiliation blocks' be placed beneath the title.
%
% NOTE: You are NOT restricted in how many 'rows' of
% "name/affiliations" may appear. We just ask that you restrict
% the number of 'columns' to three.
%
% Because of the available 'opening page real-estate'
% we ask you to refrain from putting more than six authors
% (two rows with three columns) beneath the article title.
% More than six makes the first-page appear very cluttered indeed.
%
% Use the \alignauthor commands to handle the names
% and affiliations for an 'aesthetic maximum' of six authors.
% Add names, affiliations, addresses for
% the seventh etc. author(s) as the argument for the
% \additionalauthors command.
% These 'additional authors' will be output/set for you
% without further effort on your part as the last section in
% the body of your article BEFORE References or any Appendices.

\numberofauthors{3} %  in this sample file, there are a *total*
% of EIGHT authors. SIX appear on the 'first-page' (for formatting
% reasons) and the remaining two appear in the \additionalauthors section.

\author{
% You can go ahead and credit any number of authors here,
% e.g. one 'row of three' or two rows (consisting of one row of three
% and a second row of one, two or three).
%
% The command \alignauthor (no curly braces needed) should
% precede each author name, affiliation/snail-mail address and
% e-mail address. Additionally, tag each line of
% affiliation/address with \affaddr, and tag the
% e-mail address with \email.
%
% 1st. author
\alignauthor
Rajesh Bordawekar\\
       \affaddr{IBM T. J. Watson Research Center}\\
       \affaddr{Yorktown Heights, NY 10598}\\
       \email{bordaw@us.ibm.com}
% 2nd. author
\alignauthor
Bortik Bandyopadhyay\titlenote{Work done while the author was visiting the IBM Watson Research Center.}\\
       \affaddr{The Ohio State University}\\
       \affaddr{Columbus, OH 43210}\\
       \email{bandyopadhyay.14\\@osu.edu}
% 3rd. author
\alignauthor Oded Shmueli\titlenote{Work done while the author was visiting the IBM Watson Research Center.}\\
       \affaddr{CS Department, Technion}\\
       \affaddr{Haifa 32000, Israel}\\
       \email{oshmu@cs.technion.ac.il}
}

\maketitle

\begin{abstract}
We propose Cognitive Databases, an approach for transparently
enabling Artificial Intelligence (AI) capabilities in relational
databases. A novel aspect of our design is to first view the
structured data source as meaningful unstructured text, and then use
the text to build an unsupervised neural network model using a Natural
Language Processing (NLP) technique called word embedding. This model
captures the hidden inter-/intra-column relationships between database
tokens of different types. For each database token, the model includes a vector
that encodes contextual semantic relationships. We seamlessly integrate the
word embedding model into existing SQL query infrastructure and use it
to enable a new class of SQL-based analytics queries called cognitive
intelligence (CI) queries. CI queries use the model vectors to enable complex
queries such as semantic matching, inductive reasoning queries such as analogies, predictive
queries using entities not present in a database, and, more generally,
using knowledge from external sources. We demonstrate unique
capabilities of Cognitive Databases using an Apache Spark based
prototype to execute inductive reasoning CI queries over a multi-modal
database containing text and images. We believe our
first-of-a-kind system exemplifies using AI functionality to endow relational
databases with capabilities that were previously very hard to realize in
practice. 
\end{abstract}

%   Mention word vectors: w2v, Arora
%   Mention SQL facilities (material in the document I once sent you) that use UDF and built-in synonyms
%   Explain the idea of the current paper: duality of text and structure
%   Explain the conceptual novelty over SQL (EVERYTHING IS VECTORIZED)
%   Explain exploration without schema
%   Mention options for vector sources
%   Mention maintenance
\section{Introduction}
\label{sec:intro}

\begin{displayquote}
Artificial Intelligence: Systems that perform actions that, if performed by humans, would be
considered intelligent --Marvin Minsky
\end{displayquote}

Wikipedia defines \emph{cognition} as the mental action or process of
acquiring knowledge and understanding through thought, experience, and
the senses. In broad terms,
cognition refers to the process of building knowledge capabilities
using innate resources (i.e. intelligence), enriching it with external inputs such as experiences or
interactions, and applying the knowledge to solve problems which
feeds back towards knowledge building. While these definitions are
more relevant to animate objects, they can be also applicable to
scenarios in which inanimate entities simulate cognitive processes.  

%The particular cognitive process on which
%we focus is that of reading comprehension of text  via using contexts.
%For inanimate entities such as relational databases, we start with a
%simple hypothesis: there is a large amount of untapped latent information (knowledge)
%in these systems that can be enriched and exploited to provide \textit{smart}
%querying capabilities.

We focus on a particular cognitive process of \emph{reading comprehension of
text via contexts} and apply it to \emph{relational databases}.  In the relational model, some relationships between database values and entities
 are defined at the schema level: data types, keys, and functional (and other) dependencies.
 Relationships at the instance level (i.e. actual data tables) are left
 to be explored by queries. In a strong sense, the actual semantics of the data mostly lies in
 users' minds and is expressed via queries. We take a significant diversion from this point of
 view. We postulate that there is significant \emph{latent} knowledge in a database instance irrespective
 of querying. To capture this latent knowledge we propose to use
 Artificial Intelligence (AI) techniques that take advantage of contexts.
 
Specifically, in the relational database model, the main sources of latent information
include the structure of database (e.g., column names in a relation) as well as the types of associated data
values that include unstructured natural language text, strings, numerical values,
images, SQL Dates etc. Together, these factors lead to inter- and
intra-column \emph{semantic} relationships. Current systems have  
limited support to exploit this information, namely via SQL and
extensions such as text extenders ~\cite{cutlip:db2} or RDF-based
ontologies ~\cite{lim:edbt13}.
However, SQL queries rely mainly on value-based predicates to detect
patterns. In addition, the relational data model ignores many inter-
or intra-column relationships. Thus, traditional SQL queries lack
a holistic view of the underlying relations and thus are unable to
extract and exploit semantic relationships that are collectively
generated by the various \emph{entities} in a database relation.

A few examples may serve to clarify what we mean by latent knowledge.
The first example considers a Human Resources (HR) database.
This database contains relations with information about employees,
their work history, pay grade, addresses, family members and more.  
Lately there have been some issues with an employee, John Dolittle.
As a HR professional, you are interested in names of employees who know John well.
Sure, you can get on the phone (or any other media) and start making
calls, collecting information until you have a few names of people
with whom you would like to consult. 
Much of the information you will obtain is already \emph{hidden} in the
\emph{legacy} database, but is diffused and hard to get at.  
This may include people who worked with John, managed him,
complimented him, complained about him, provided technical services to
him, were members of a small team with him and so on.   
Wouldn't it be nice if you could write a SQL query that would 
use this hidden knowledge and essentially ask \emph{provide the names of the 10 employees most related to
John Dolittle}. 

%In the previous example, the information  was diffused across the database.
%Next, we look at a more focused type of problem.
%You are serving as a journal editor.
%Suppose you have a database of papers, authors and venues (perhaps derived from DBLP).
%You need to find referees for a paper, say \emph{'First Steps in Endowing
%Databases with Intelligence'}. You may wonder who, based on their
%previous work, is most likely to be a good referee for this
%paper. Again, you can start asking around, browsing and eventually
%come up with a good set of referees. It would be convenient however to
%just pose a SQL query expressing \emph{who are the most likely authors of a
%paper entitled 'First Steps in Endowing Databases with
%Intelligence'}. Presumably, from the result, you'll be able to
%identify the best suited referees. 

In the previous example, we relied on the content of the database in isolation. 
The next example involves entities external
to the database. Suppose you have  database of active vacations, 
featuring diving, hiking, skiing, desert driving and more.
You are a bit worried as to which vacation package is the most dangerous one.
Naturally, the official descriptions in the database will not always provide the information.
It is likely that the words \emph{accident}, \emph{danger}, \emph{wounded}, and
\emph{death} will not even be present in the database. These words are
present in other data sources, from Wikipedia to news
articles. Suppose you have access to such external sources. 
Wouldn't it be nice if you could utilize these external sources and
pose a SQL query expressing \emph{which are the most dangerous vacation
packages}. In this example, as well as the previous ones, you may want also to
have a degree of certainty associated with each potential answer. 

It is worth pointing out what distinguishes the level of intelligence we are 
looking for from known extensions to
relational systems. In current systems one needs to pose a query based on 
some knowledge of the relational schema.
The query may be \emph{assisted} by text-aware features such as  DB2 Text
Extender~\cite{cutlip:db2}, WordNet or using RDF-based ontologies~\cite{lim:edbt13}.
These may be used to identify synonyms and related terms and relax the
query by allowing it to explore more possibilities than those
explicitly specified by the user~\cite{chu:relax}. Of course, such relaxation may result in obtaining a larger result set.
But, all these useful features assume that the user knows how to specify a backbone query.
The example problems we listed above are such that formulating an effective SQL
query is a daunting task. In fact, these examples resemble research
projects rather than standard queries. One can also allow the user to
specify the query in natural language~\cite{Li:nlq}, but this pushes the problem of
expressing a query to an automated tool; again, it is unclear how a
tool will approach these problem if the tool's writer does not have a
ready recipe. This highlights the need of a new set of tools to enable
far richer querying. 

In this paper, we explore the potential of using Natural Language Processing (NLP) approaches to endow databases with query expression capabilities that
were very hard, or perhaps impossible, to realize in practice, and at a reasonable cost in terms of storage overhead as well as
processing time. The unique aspect of our proposal is to 
first represent the data and optionally, schema, of a (structured) relational database
as an unstructured text document and then use a NLP technique,
\emph{Vector Space Models (VSM)}~\cite{salton:vsm}, to extract
latent semantic relationships via associations in the generated text. The trained
VSM model represents the semantic
meaning of the words as vectors and enables operations on these
vectors to mimic cognitive operations on natural language words. As
these words represent relational entities and values, the VSM model,
in fact, captures intra-/inter-column relationships in the
relational database. We then integrate the VSM model into an existing standard SQL query
processing system and expose the novel vector-based cognitive
operations via a new class of SQL analytics queries, called
\emph{Cognitive Intelligence (CI)} queries~\cite{bordawekar:corr-abs-1603-07185}. We believe this is one of
the first examples of AI \emph{transperently augmenting} a relational
database system. Clearly, this is only one of the many possible ways
of integrating AI capabilities in database systems, e.g., for enhancing their querying
capabilities or improving their operational
capabilities~\cite{Park:sigmod17, VanAken:sigmod17, Li:nlq}.  While our
current focus is on enhancing relational databases, we believe this
approach can be applied to other database domains such as XML/RDF or
JSON databases, document databases, graph databases, and key-value stores.

We are currently developing an Apache Spark-based prototype to implement our
vision of an AI-enhanced cognitive relational database. The rest of the paper
provides more details on the design and implementation of such a system. In Section~\ref{sec:design}, we introduce the vector space
modeling process and detail the execution flow of the system we envisage. In
Section \ref{sec:model}, we provide specifics of the data preparation
and building a specialized vector space model. In Section \ref{sec:system} we discuss
three significant classes of CI queries:
Similarity Queries, Inductive Reasoning Queries and
Cognitive OLAP Queries; we also present cognitive extensions to the
Relational Data Model. We also describe the design of cognitive User
Defined Functions (UDFs). Section~\ref{sec:practice} describes a
practical scenario which demonstrates unique aspects of a cognitive
database system: ability to invoke inductive reasoning (e.g.,
analogies, semantic clustering, etc.) queries over multi-modal data (e.g., images and text). We also discuss CI query performance issues, with
focus on an important building block:  Nearest-Neighbor Computations.
Related work is discussed in Section \ref{sec:related} and we conclude
by outlining extensions, future work, and success criteria in Section \ref{sec:concl}.

\section{Designing a Cognitive Database}
\label{sec:design}

Our goal is to build a cognitive relational database system that not
only extracts latent semantic information, but can also enrich it by using
external input (e.g., external knowledge bases, new data being
inserted, or types of invoked queries) and use it \emph{transparently} to
enhance its query capabilities. To achieve these goals, we rely on the
VSM approach that infers word meanings using the
distributional hypothesis, which states that words in a neighborhood
(or context) contribute to each other's
meanings~\cite{harris:distr,hinton:distr}. Specifically, we use a
\emph{predictive}~\cite{baroni:compare} implementation of the VSM approach, commonly referred to as \emph{word
  embedding},  which assumes a probabilistic language model to capture
relationships between neighborhood words~\cite{bengio:jmlr03}. 
The word embedding approach fixes a $d$-dimensional vector space
and associates a vector of continuous-valued
real numbers to a word to encode the meaning of that word. Thus, for a given text corpus, the 
meaning of a word reflects collective contributions of neighborhood
words for different appearances of the word in the corpus. Two words
are closely related or have similar meaning if they appear often
within close proximity of the same, or similar meaning, words. If two words have similar
meaning, their meaning vectors point in very similar directions, i.e.,
the cosine similarity between their vectors is high (vector similarity
is measured as a cosine of the angle between two vectors and can vary
from 1.0 to -1.0).  One surprising application of word-embedding
vectors is their usage in solving inductive reasoning problems such
as computing analogies by using vector algebra calculations~\cite{Sternberg1979, Rumelhart1973, levy:conll14}.

Over a past few years, a number of methods have
been developed to implement the word embedding. Recently, an
\emph{unsupervised} neural network based approach,
\newline Word2Vec~\cite{w2v,mikolov:nips13}, has gained popularity due to its
performance and ability to capture syntactic as well semantic
properties of words. We use Word2Vec as it is
easy to adapt, can be trained incrementally, and can be used for
building models from both structured and unstructured data sources
(alternatively, we could use approaches such as GloVe~\cite{pennington:glove14}).

\begin{figure}[htbp]
  \centering
                \includegraphics[height=1.2in]{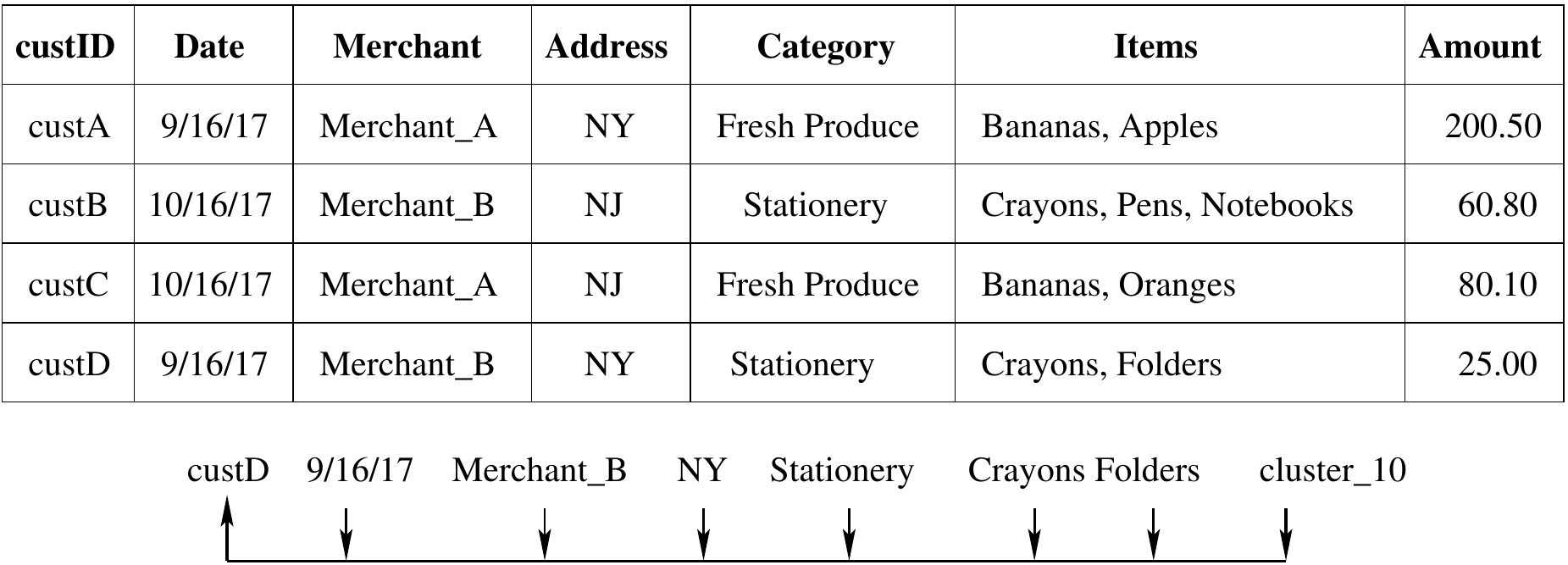}
                \caption{Example of customer analytics}
\label{fig:cust1}
\end{figure}

\begin{figure}[htbp]
  \centering
                \includegraphics[height=1.2in]{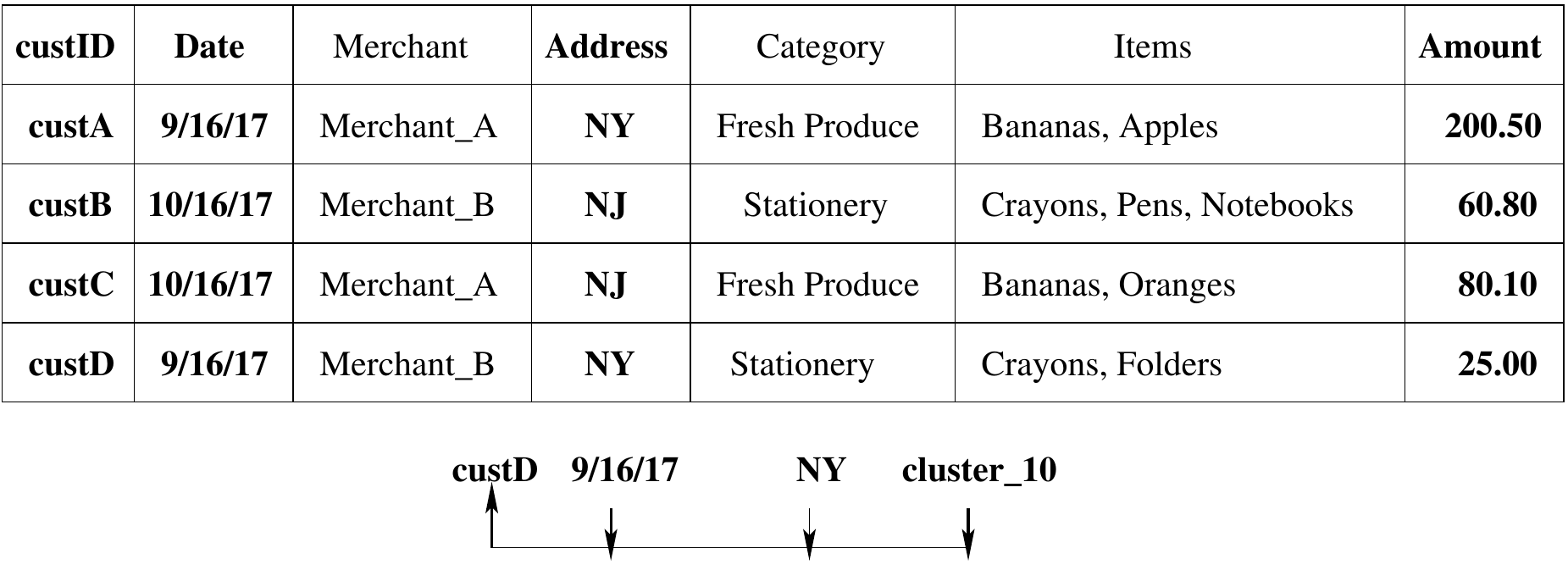}
                \caption{Example of customer analytics with a different
                  relational view}
\label{fig:cust2}
\end{figure}

In the database context, vectors may be produced by either learning
on text transformed and extracted from the database itself and/or using external text sources. For learning from
a database, a natural way of generating vectors is
to apply the word embedding method to a string of tokens generated from the database: each row 
(tuple) would correspond
to a sentence and a relation would correspond to a document. Thus, vectors enable a
\emph{dual view} of the data: relational and meaningful text. To
illustrate this process, consider  Figures~\ref{fig:cust1} and~\ref{fig:cust2} that present a simple customer sales table.
Figure~\ref{fig:cust1} shows an English sentence-like representation
of the fourth row in the table (note that the numeric value
\texttt{25.00} is represented by the string \texttt{cluster\_10}. We
discuss the reasons in the next section.). Using the scope of the
generated sentence as the context, the word embedding approach
infers latent semantic information in terms of token associations and
co-occurrences and encode it in vectors. Thus, the vectors
capture first inter- and intra-column relationships within a row
(sentence) and then aggregate these relationships across the relation (document) to
compute the \emph{collective semantic} relationships. At the end of training, each
unique token in the database would be associated with a
$d$-dimensional meaning vector, which can be then used to query the
source database. As a simple example, the relational entity
\texttt{custD} is \emph{semantically} similar to \texttt{custB} due to
many common semantic contributors (e.g., \texttt{Merchant\_B},
\texttt{Stationery}, and \texttt{Crayons}). Equivalently,
\texttt{custA} is similar to \texttt{custC} due to similar reasons.

We may use a relational view of a table, rather than the original
table, to generate text representing the database content. 
This may be useful for a supporting a particular class of applications.
Consider a scenario in which a view of the table is defined
(Figure~\ref{fig:cust2}), the  view only projects data from the bold columns (\texttt{Cust}, \texttt{Date}, \texttt{Address} and \texttt{Amount}). In this
case, the generated sentence-like
representation would be different than the first case. Hence, it
will generate a different word embedding model.
%and results in a
%different semantic clustering: now \texttt{custD} will be similar to
%\texttt{custA}, and \texttt{custB} will be similar to \texttt{custC}.

This examples illustrates a key design feature of our cognitive
database: the neighborhood context used for building the word
embedding model is determined by the relational view being
used. Hence, \emph{the inferred semantic meaning of the relational
entities reflect the collective relationships defined by the
associated relational view.}

\begin{figure}[htbp]
  \centering
                \includegraphics[height=1.25in]{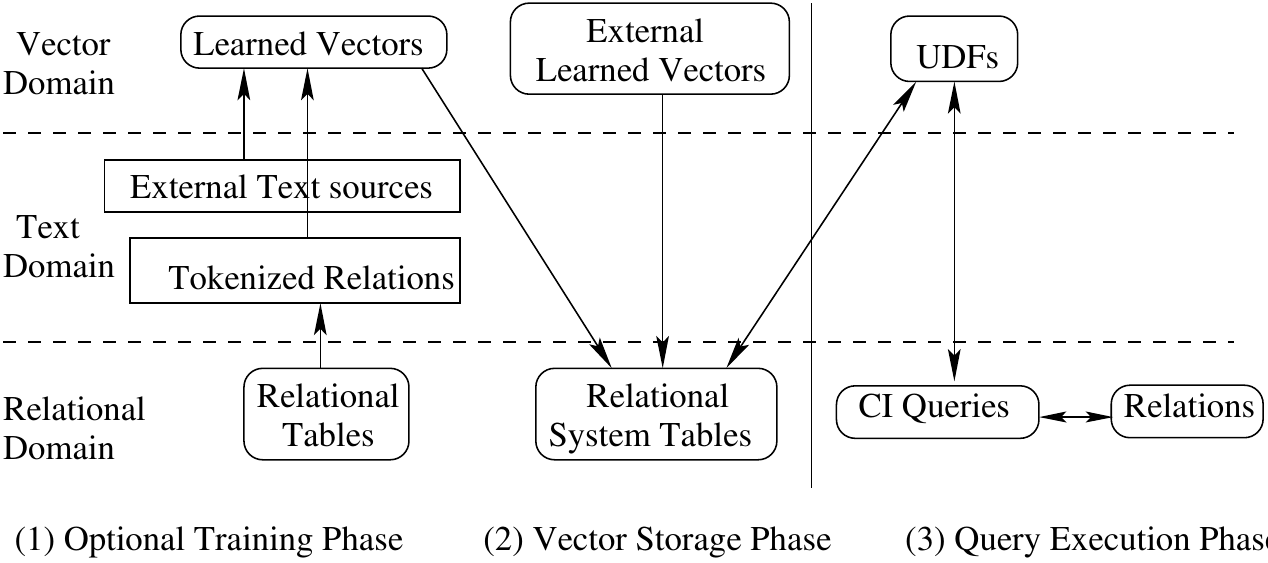}
                \caption{End-to-end execution flow of a cognitive relational database}
        %\vskip -14pt
\label{fig:flow}
\end{figure}

The cognitive relational database has been designed as an
\emph{extension} to the underlying relational database,
and thus supports all existing relational features. The cognitive
relational database supports a new class of business intelligence (BI)
queries called \emph{Cognitive Intelligence} (CI) queries. The CI
queries extract information from a relational database based, in part,
on the contextual semantic relationships among database entities, encoded as meaning vectors.  
Figure~\ref{fig:flow} presents key phases in the end-to-end execution flow of a
cognitive relational database system.  The first, optional, phase
involves (1) Generating token sequences from the database tables
(\emph{textification}), and then applying a word embedding model
training method on the unstructured text corpus created from these
token sequences. Following model training, the resultant
vectors are stored in a relational system table (phase 2). At runtime,
the SQL query execution engine uses various user-defined functions (UDFs) that fetch the
trained vectors from the system table as needed and answer CI queries
(phase 3). The CI queries take relations as input and return a relation
as output. CI queries \emph{augment} the capabilities of the
traditional relational BI queries and can be used in conjuction with
existing SQL operators (e.g., OLAP~\cite{gray:olap}).

\section{Building the Semantic Model}
\label{sec:model}

\begin{displayquote}
The key to artificial intelligence has always been the
representation. --Jeff Hawkins
\end{displayquote}

In this section, we discuss how we train a word embedding model using
data from a relational database. Our training approach is
characterized by two unique aspects: (1) Using \emph{unstructured
  text} representation of the structured relational data as  input to
the training process (i.e. irrespective of the associated SQL types, all entries from a relational database are
converted to  text tokens representing them), and (2) Using the
\emph{unsupervised} word embedding technique to generate meaning
vectors from the input text corpus. Every unique token from the
input corpus is associated with a meaning vector. We now elaborate on these
two aspects.

\subsection{Data Preparation}
\label{sec:model-prep}

\begin{figure}[htbp]
  \centering 
  \includegraphics[height=1.35in]{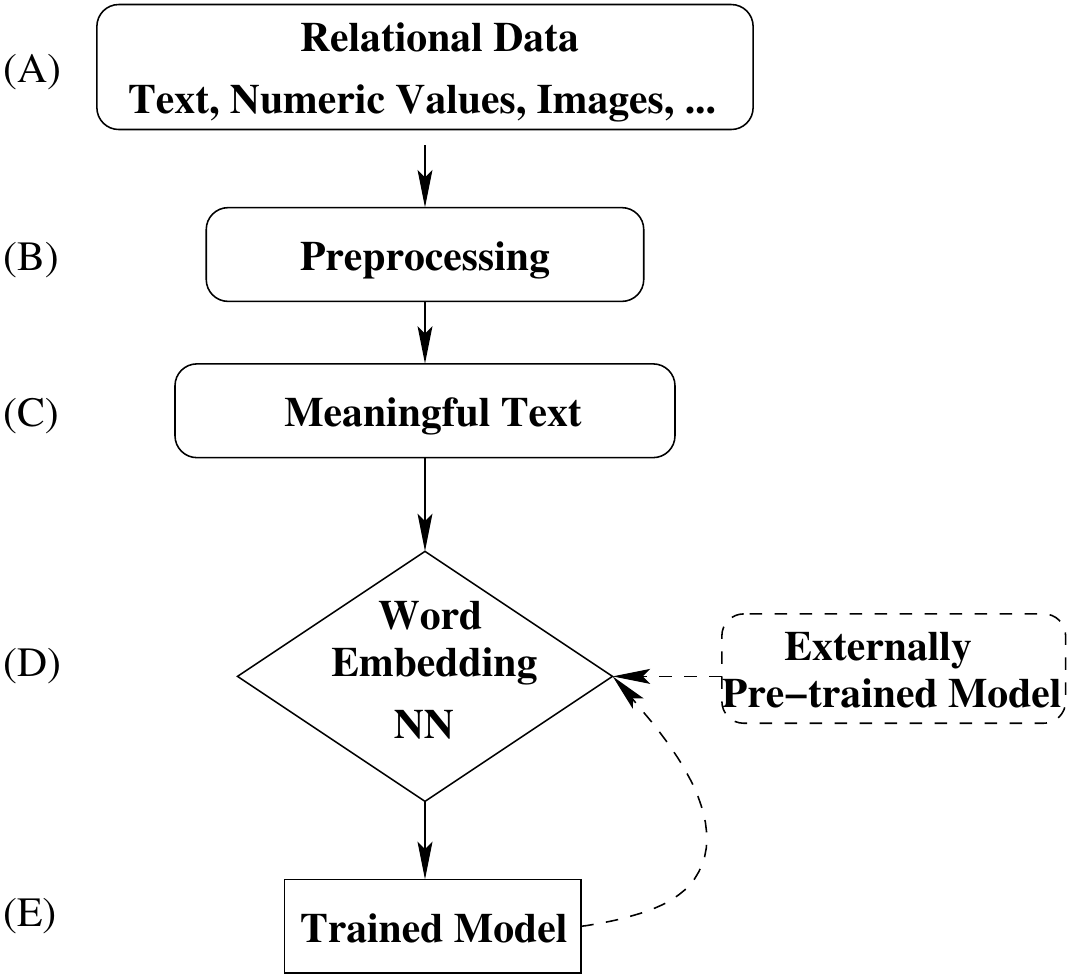}
  \caption{Multiple stages in creating the word embedding model}
\label{fig:prepare}
\end{figure}

The data preparation stage takes a relational table with different SQL
types as input and returns an unstructured but meaningful text corpus consisting of a set
of sentences. This transformation allows us to generate a uniform semantic
representation of  different SQL types. This process of
\emph{textification} requires two stages: data pre-processing and text
conversion (Figure~\ref{fig:prepare}).

The textification phase processes each relational row seperately and
for each row, converts data of different SQL data types to text. In some scenarios, one may want to build a model
that also captures relational column names. For such cases, the
pre-processing stage first processes the column names before
processing the corresponding data.

For SQL variables of \texttt{VARCHAR} type, preprocessing involves one
or more of the following actions: (1) prepend the column attribute
string to a SQL variable, (2) creating a single \emph{concept} token
from a group of \texttt{VARCHAR} tokens, e.g., \texttt{JPMorgan Chase} is represented as
\texttt{JPMorgan\_Chase}, (3) creating a single \emph{token} for
semantically similar sequences of \texttt{VARCHAR} tokens, e.g., two 
sequences of tokens, \texttt{bank of america} and \texttt{BANK OF
  AMERICA}, can be represented by a single compound token 
\texttt{BANK\_OF\_AMERICA}, and (4) Using an external mapping or
domain-specific ontologies to create a common representative token for
a group of different input tokens. This approach is useful for
enabling \emph{transfer learning} via reusing the same training model
for a group of related tokens. After pre-processing, all input text
tokens have uniform representations.

In addition to text tokens, our current implementation supports
numeric values and images (we assume that the database being queried
contains a \texttt{VARCHAR} column storing links to the images). These
techniques can be applied to other SQL datatypes such as SQL Date as
well. For
numeric values, we use three different approaches to generate
equivalent text representations: (1) creating a string version of the
numerical value, e.g.,  value \texttt{100.0} for the column name
\texttt{price} can be represented by either \texttt{PRICE\_100.0}
or \texttt{``100.0''}, (2) User-managed categorization: a user can
specify rules to define ranges for the numeric values and use them to
generate string tokens for the numeric values. For example, consider
values for a column name, \texttt{Cocoa Contents}. The value
\texttt{80\%}, can be replaced by the string token
\texttt{choc\_dark}, while the value \texttt{35\%}, can be replaced by
the string token \texttt{choc\_med}, etc., and (3) user-directed
clustering: an user can choose values of one or more numerical columns
and cluster them using traditional clustering algorithms such as
K-Means. Each numeric value is then replaced by a string representing
the cluster in which that value lies (e.g., \texttt{cluster\_10}  for
value \texttt{25} in Figure 1). 

%It should be noted that creating
%unique string for a numerical value is equivalent to assigning an
%unique numeric value to its own cluster, and using the string
%representation of the cluster.

For image data, we use approaches similar to ones used for numerical
values. The first approach represents an image by its string token, e.g., a string representing
the image path or a unique identifier. The second approach uses
pre-existing classifers to cluster images into groups and then uses
the cluster information as the string representation of the image. For
example, one can use a domain-specific deep neural network (DNN) based
classifier to cluster input images into classes~\cite{jia2014caffe} and then use the
corresponding class information to create the string identifiers for
the images. The final approach applies of-the-shelf image to tag
generators, e.g., IBM Watson Visual Recognition System (VRS) ~\cite{VRS}, to
extract image features and uses them as string identifiers for an
image. For example, a Lion image can be represented by the following
string features, \texttt{Animal}, \texttt{Mammal}, \texttt{Carnivore},
\texttt{BigCat}, \texttt{Yellow}, etc.

Once text, numeric values and images are replaced by their text
representations, a relational table can be viewed as unstructured
meaningful text corpus to be used for building an word
embedding model. For \texttt{Null} values of these types, we replace them by
the string \texttt{column\_name\_Null}. The methods outlined here can be applied to other
data types such as SQL \texttt{Date} and spatial data types such as lattitude and longitude.

\subsection{Model Training}

We use an \textbf{unsupervised} approach, based on the Word2Vec (W2V)
implementation~\cite{w2v}, to build the word embedding model from the relational
database data. Our training approach operates on the unstructured text
corpus, organized as a collection of \emph{English-like} sentences,
separated by stop words (e.g., newline).  There is no need of
labelling the training data as we use unsupervised training. Another
advantage of unsupervised training is that users do not need to do
any feature engineering~\cite{goodfellow-deep}; features of the training set are extracted
automatically by the training process.

During model training, the classical W2V implementation uses a
simplified 3-layer shallow neural network that views the input text corpus as a
sequence of sentences. For each word in a sentence, the W2V code defines a neighborhood
window to compute the contributions of nearby words. Unlike deep learning based classifiers, the
output of W2V is a set of vectors of real values of dimension $d$, one
for each unique token in the training set (the vector space dimension
$d$ is independent of the token vocabulary size).  In our scenario, a
text token in a training set can represent either text, numeric, or
image data.  Thus, the model builds a \emph{joint} latent representation that integrates information
across different \emph{modalities} using \emph{untyped uniform}  feature (or \emph{meaning})
vectors. 

Our training implementation builds on the classical W2V
implementation, but it varies from the classical approach in a number
of ways (Figure~\ref{fig:prepare}): 

\begin{itemize}

 \item A sentence generated from a relational row is generally
not in any natural language such as English.\footnote{Currently, we assume that database tokens are
  specified using the English language.} Therefore,  W2V's assumption
that the influence of any word on a
nearby word decreases as the word distances increases, is not
applicable. In our implementation, every token in the training set has
the same influence on the nearby tokens; i.e. we view the generated sentence as
a bag of words, rather than an ordered sequence.

\item Another consequence is that unlike an
  English sentence, the last word is equally related to the first word
  as to its other neighbors. To enable such relationships, we use a
  circular neighborhood window that wraps around a sentence (i.e. for the
  last word, the first word can be viewed as its immediate neighbor).

\item For relational data, we provide special consideration to
  primary keys. First, the classical W2V discards less frequent words
  from computations. In our implementation, every token, irrespective
  of its frequency, is assigned a vector. Second, irrespective of the
  distance, a primary key is considered a neighbor of every other word
  in a sentence and included in the neighborhood window for each word. Also, the neighborhood
  extends via foreign key occurrences of a key value to the row in
  which that value is key.

\item In some cases, one may want to build a model in which values of particular
  columns are given higher weightage for their contributions towards
  meanings of neighborhood words. Our implementation enables users to
  specify different weights (or attention~\cite{bahdanau:attention})
  for different columns during model training (in this scenario, one needs to use
  a training set in which column names are  embedded).

\item Finally, our implementation is designed to enable
  \emph{incremental} training, i.e. the training system takes as input a
  pre-trained model and a new set of generated sentences, and returns an
  updated model. This capability is critical as a database can be
  updated regularly and one can not rebuild the model from scratch
  every time. The pre-trained model can be built from the database
  being queried, or from an external source. Such sources may be
  publicly available general sources (e.g., Wikipedia), text from a
  specific domain (e.g., from the FDA regarding medical drugs), text textified
  from other databases or text formed from a different subset of
  tables of the same database. The use of pre-trained
  models is an example of \emph{transfer} learning, where a model
  trained on an external knowledge base can be used either for querying
  purposes or as a basis of a new model~\cite{goodfellow-deep}.

\end{itemize}

\begin{figure}
   \centering
   \includegraphics[height=1.5in]{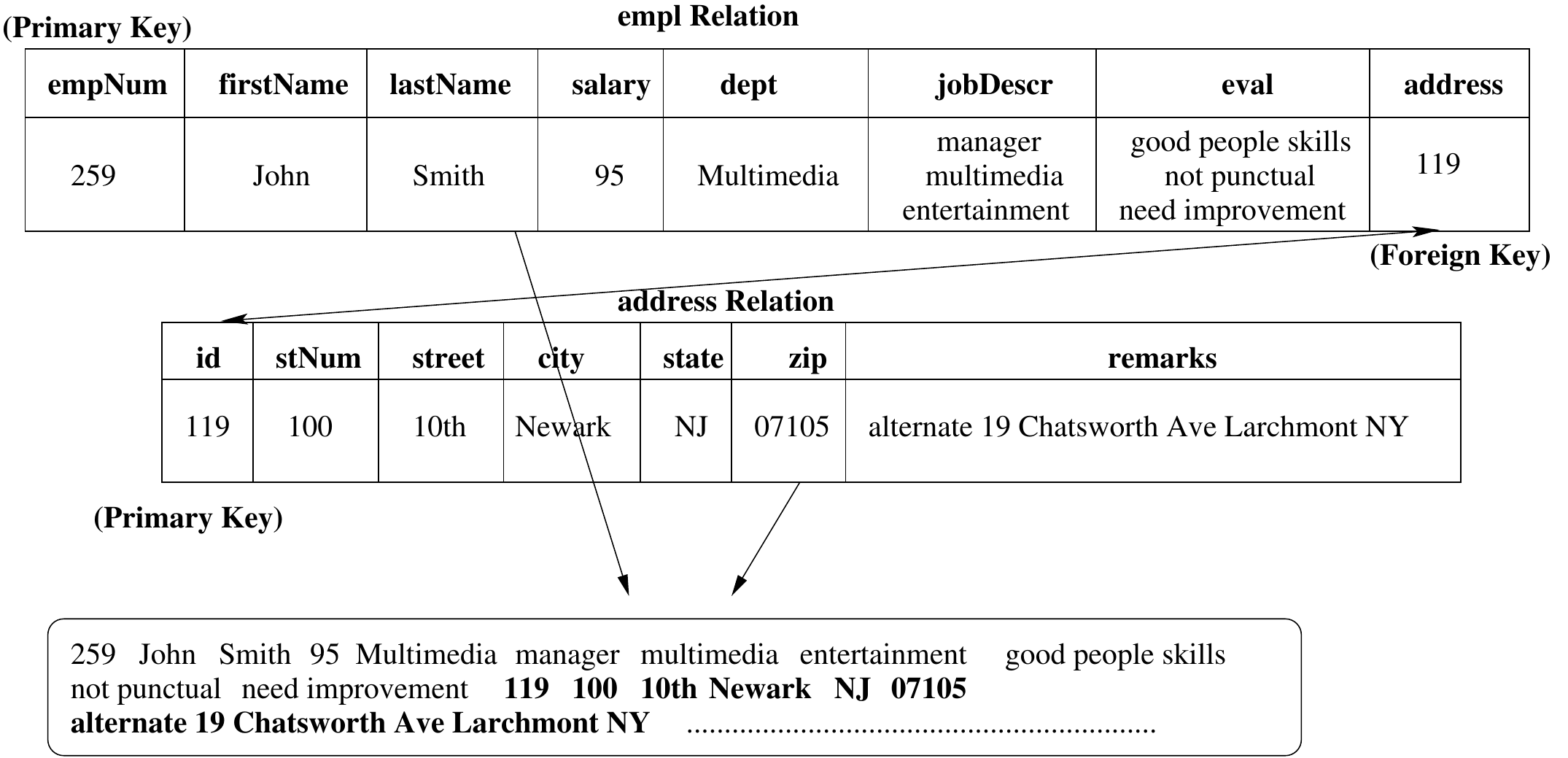}
   \caption{Text view of two tables joined using primary and foreign keys}
\label{fig:foreign}
\end{figure}

In practice, enterprise database systems, as well as data warehouses,
are built using many inter-related database tables. Forming a training
corpus from multiple tables is non-trivial.  There are numerous
options, including: 
\begin{itemize}  
\item Build separate models (i.e. a set of word vectors), each based
  on an individual, informative, table.

\item Build models each based on linked tables where, usually, the
  linking is based on foreign keys appearing in say table \texttt{A} pointing
  to tuples into another, say table \texttt{B}. When a foreign key is present, 
  during tokenization of table \texttt{A}, we can follow the foreign key to a 
  row in table \texttt{B}. We can then tokenize fields of interest in
  the row of table \texttt{B} and insert the resulting sequences into the
  sequence generated for table \texttt{A}. Figure~\ref{fig:foreign} presents another
  example of a database table, \texttt{address}, and a resulting token sequence
  that utilizes a relationship between the \texttt{empl} table and the
  \texttt{address} table; namely the \texttt{address} table provides the addresses
  for the employees of database table \texttt{empl}. 
 Technically, the
  resulting token sequence is based on foreign key \texttt{119}
  in the \texttt{address} column of the table \emph{emp} which 
  provides a value for key column \texttt{id} of the \texttt{address} table. The
  straight forward way to tokenize with foreign keys is to insert the
  subsequence generated out of the \texttt{B} row immediately after the one
  generated for the \texttt{A} row as depicted in Figure~\ref{fig:foreign}; another possibility
  is to intermix the subsequence from the \texttt{B} row within the \texttt{A} row
  sequence following the tokenization of the foreign keys values (again, other options may apply). 

\item A collection of tables may be identified and textified into a
  collection of texts. These texts may be concatenated to form a
  single text which may be used in training. The tables in this
  collection should form a coherent informative subset of the
  database. 
\item In all the options above, the training text may be augmented
  with text from external sources.  
\end{itemize}

\section{Building a Cognitive Database System}
\label{sec:system}

Cognitive intelligence (CI) queries are standard SQL queries and can be
implemented using the existing SQL query execution infrastructure. The distinguishing
aspect of cognitive intelligence queries, contextual semantic comparison between
relational variables, is implemented using user-defined functions
(UDFs). These UDFs, termed \emph{cognitive} UDFs, take
\emph{typed} relational values as input and compute semantic relationships between them using
\emph{uniformly untyped} meaning vectors. This enables the relational
database system to seamlessly analyze data of different types (e.g., text,
numeric values, and images) using the same  SQL CI query.  

Our current implementation is built on the Apache Spark 2.2.0
infrastructure. Our system follows the cognitive database execution
flow as presented in Figure~\ref{fig:flow}. The system first initializes in-memory Spark Dataframes from
external data sources (e.g., relational database or CSV files), loads
the associated word embedding model into another Spark Dataframe (which can be created offline from
either the database being queried or external knowledge bases such as Wikipedia),
and then invokes CI queries using Spark SQL. The SQL queries invoke
Scala-based cognitive UDFs to enable computations on
the meaning vectors (we also provide a Python based implementation).

\subsection{Design of Cognitive UDFs} 

A cognitive UDF takes as input either relational query variables or
constant tokens, and returns a numeric similarity value that measures the semantic relationships between the input
parameters. A user can then control the result
of the CI query by using a numerical bound for similarity result value
as a predicate for selecting eligible rows. A user can also use
SQL ordering clauses, \texttt{DESC} or \texttt{ASC}, to order results
based on the similarity value that captures the semantic closeness between the relational variables:
higher is the similarity value, closer are these two relational 
variables. The UDFs perform three key tasks: (1) processing
input relational variables to generate tokens used for training. This
involves potentially repeating the steps executed during the data
preparation stage, such as creating compound tokens. For numeric values, one can use the centroid
information to identify the corresponding clusters. For images, the
UDF uses the image name to obtain corresponding text tokens, (2) Once
the training tokens are extracted, the UDF uses them to fetch corresponding
meaning vectors, and (3) Finally, the UDF uses the fetched vectors to execute
similarity computations to generate the final semantic relationship
score.  

The basic cognitive UDF operates on a pair of sets (or sequences) of tokens
associated with the input relational parameters (note: value of a
relational parameter can be a set, e.g., \{\texttt{Bananas, Apples}\}, see
Figure~\ref{fig:cust1}).  The core computational operation of a 
cognitive UDF is to calculate similarity between a pair of tokens by
computing the cosine distance between the corresponding vectors. For
two vectors $v_1$ and $v_2$, the cosine distance is computed as
$cos(v_1,v_2)=\frac{v_1\cdot v_2}{\lVert {v_1} \rVert \lVert
  {v_2} \rVert}$. The cosine distance value varies from 1.0 (very
similar) to -1.0 (very dissimilar). For sets and sequences, the individual pair-wise similarity values are
then aggragated to generate the result. In case of sequences,
computation of the  final similarity value takes into account the
\emph{ordering} of tokens: different pair-wise distances contribute differently
to the  final value based on their relative ordering. For example, two
food items, a chicken\_item consisting of (chicken, salt), is not very
similar to a corn\_item consisting of (corn, salt), although both
contain salt (however, the corn\_item is closer to chicken\_item than
a wheat\_item that contains (wheat, sugar)). 

\subsection{Cognitive Intelligence Queries}
\label{sec:queries}

The basic UDF and its extensions are invoked by the SQL CI queries to
enable semantic operations on relational variables. Each CI query uses
the UDFs to execute \emph{nearest neighbor} computations using the vectors from the
current word-embedding model.  Thus, CI queries provide
\emph{approximate} answers that \emph{reflect}  a given
model. The CI queries can be broadly classified into four categories as follows:

\noindent \textbf{(1) Similarity/Dissimilarity Queries:} The basic UDF
that compares two sets of relational variables can be integrated into
an existing SQL query to form a \emph{similarity} CI
query. Figure~\ref{fig:similarity-sql} illustrates a SQL CI query
that identifies similar customers by comparing their
purchases. Assume that \texttt{sales} is a table that contains all
customer transactions for a credit card company and whose
\texttt{sales.Items} column contains all items 
purchased in a transaction (Figure~\ref{fig:cust1}). The current query uses a UDF,
\texttt{similarityUDF()}, that computes similarity match between two sets of
vectors, that correspond to the items purchased by the corresponding
customers. Unlike the food item scenario, the purchased item list can
be viewed as an unordered bag of items; and individual pair-wise
distances contribute equally to the final result. The query shown in
Figure~\ref{fig:similarity-sql}  uses the similarity score to select
rows with related customers and returns an ordered set of similar customer
IDs sorted in descending order of their similarity score. This
query can be easily tweaked to identify dissimilar customers based on
their purchases. The modified CI query will first choose rows whose
purchases have lower similarity (e.g., $< 0.3$) and if the results
are ordered in an ascending form using the SQL \texttt{ASC} keyword, returns customers
that are highly dissimilar to a given customer (i.e., purchasing
very different items). If the results are ordered in the
descending order using the SQL \texttt{DESC} keyword, the CI query
will return customers that are somewhat dissimilar to a given
customer. 

\begin{figure}[htbp]
\hrule
{\small \tt{
\begin{flushleft}
\ix \\
SELECT X.custID, Y.custID,  similarityUDF(X.Items, Y.Items) AS similarity\\
FROM sales X, sales Y\\
WHERE similarityUDF(X.Items, Y.Items) > 0.5\\
ORDER BY similarity DESC\\
\end{flushleft}
}
\hrule
}
\caption{Example of an CI similarity query: find similar customers based
  on their purchased items}
\label{fig:similarity-sql}
\end{figure}

A modified version of the query (not shown) can identify similar
customers based on their overall purchasing pattern as 
evidenced in a number of rows. The word embedding model  creates a
vector for each customer name that captures the overall purchases made
by that customer. Then, the customers with similar purchase patterns
would have vectors that are close using the cosine distance
metric. The pattern observed in this query can be applied to other
domains as well, e.g., identifying patients that are taking similar
drugs, but with different brand names, or identifying food items with
similar ingredients, or  recommending mutual funds with similar investment
strategies. As we will see in the next section, the similarity query
can be applied other data types, such as images.

\begin{figure}[htbp]
\hrule
{\small \tt{
\begin{flushleft}
\ix \\
SELECT X.custID, Y.custID,  Y.Merchant, valueSimUDF(X.Amount, Y.Amount) AS similarity\\
FROM sales X, sales Y\\
WHERE X.custID='custA' AND valueSimUDF(X.Amount, Y.Amount) > 0.5 AND
X.custID != Y.custID AND X.amount > 150.0 AND Y.amount < 100.0\\
ORDER BY similarity DESC\\
\end{flushleft}
}
\hrule
}
\caption{Example of an CI value similarity query: find similar
  transactions using purchased amount for comparison}
\label{fig:valSim-sql}
\end{figure}

Figure~\ref{fig:valSim-sql} presents a CI query executing similarity
operations using a numeric variable. For the sake of example, assume
that no two transactions have the same amount value and each unique
numerical value is associated with its own string token. In this scenario, one wants
to identify transactions from the \texttt{sales} table
(Figure~\ref{fig:cust1}): using similarity based on the purchase
amount (in this case, \texttt{200.50} for customer \texttt{custA}). The
UDF, \texttt{valueSimUDF()}, takes two numeric values as input
parameters, and compares them using their overall \emph{context} (which is
captured in their meaning vectors), not their numerical values. The most
similar amount to \texttt{200.50}, would be \texttt{80.10} (for customer \texttt{custC}), as it shares the
most context (e.g., category, address, merchant, and items). The
least similar amount would be \texttt{60.80} as it has completely
different context than amount \texttt{200.50}. This example also illustrates 
how one can combine the value-based and semantic-based comparisons in 
the same SQL query.

\begin{figure}
\hrule
{\small \tt{
\begin{flushleft}
\ix \\
SELECT X.custID, similarityUDF(X.Items, `listeria') AS similarity\\
FROM sales X\\
WHERE similarityUDF(X.Items, `listeria') > 0.3\\
ORDER BY similarity DESC\\
LIMIT 10\\
\end{flushleft}
}
\hrule
}
\caption{Example of a prediction query: find customers that have
  purchased items affected by a listeria recall}
\label{fig:allergic-sql}
\end{figure}

The third use case provides an illustration of a \emph{prediction} CI
query  which uses a model that is externally trained using an
unstructured data source or another database
(Figure~\ref{fig:allergic-sql}). Consider a scenario of a recall of
various fresh fruit types due to possible listeria infection. This example
assumes that we have built a word embedding model using the recall
notices as an external source.  Assume that the recall document
lists all fruits impacted by the possible listeria infection, e.g.,
\emph{Apples, Peaches, Plums, Nectarines,..}. The model will create vectors for all
these words and the vector for the word \emph{listeria} will be closer
to the vectors of \emph{Apples}, \emph{Peaches}, \emph{Plums}, etc. Now, we can import this model and use it to query
the \texttt{sales} database to find out which customers have bought items that
may be affected by this recall, as defined by the external source. As
Figure~\ref{fig:allergic-sql} shows, the \texttt{similarityUDF()} UDF is
used to identify those purchases that contain items similar to
\emph{listeria}, such as  \emph{Apples}. This example
demonstrates a very powerful ability of CI queries that enables users
to query a database using a token \textbf{not present}
in the database (e.g., \emph{listeria}). This capability can be applied to different scenarios
in which recent, updatable information, can be used to query
historical data. For example, a model built using a FDA recall
notices could be used to identify those customers who have purchased
medicines similar to the recalled medicines.

 \noindent \textbf{(2) Inductive Reasoning Queries:} An unique
 feature of word-embedding vectors is their capability
 to answer \emph{inductive reasoning} queries that enable an individual to
 reason from \emph{part} to \emph{whole}, or from \emph{particular} to
 \emph{general}~\cite{Rumelhart1973, Sternberg1979}. Solutions to
 inductive reasoning queries exploit latent semantic structure in the 
 trained model via algebraic operations on the corresponding
 vectors. We encapsulate these operations in UDFs to support following five types of inductive reasoning queries:
 analogies, semantic clustering, analogy sequences, clustered
 analogies, and odd-man-out~\cite{Rumelhart1973}.

\begin{itemize}

\item Analogies: Wikipedia defines analogy as a process of
  transferring \emph{information} or \emph{meaning} from one subject
  to another. A common way of expressing an analogy is to use
  relationship between a pair of entities, \texttt{source\_1} and
  \texttt{target\_1}, to reason about a possible target entity,
  \texttt{target\_2}, associated with another known source entity,
  \texttt{source\_2}. An example of an analogy query is $Lawyer:Client
  :: Doctor:?$, whose answer is $Patient$.  To solve an analogy problem of the form
$(X:Y::Q:?)$, one needs to find a token $W$ whose meaning 
vector, $V_w$, is closest to the ideal response vector $V_R$, where $V_R=(V_Q+V_Y-V_x)$
\cite{Rumelhart1973}. Recently, several solutions have
been proposed to solve this formulation of the analogy query ~\cite{levy:conll14,
  linzen2016issues, mikolov:nips13}. We have implemented the 3COSMUL
approach~\cite{levy:conll14} which uses both the absolute distance and
direction for identifying the vector $V_W$ as 

\begin{equation}\label{3cosmul}
\argmax_{W \in C} \frac{cos(V_W, V_Q) cos(V_W, V_Y)}{cos(V_W, V_X) + \epsilon}
\end{equation}

where $\epsilon=0.001$ is used to avoid the denominator becoming
0. Also, 3COSMUL converts the cosine similarity value of $c$ to
$\frac{(c+1)}{2}$ to ensure that the value being maximized is
non-negative. 

Figure~\ref{fig:analogy-sql} illustrates a CI query that
performs an analogy computation on the relational variables using the
UDF \texttt{analogyUDF()}. This query aims to find a customer from the
\texttt{sales} table (Figure~\ref{fig:cust1}), whose relationship to the category, \texttt{Fresh
  Produce}, is similar to what \texttt{C3423567} has with the category,
\texttt{Frozen Goods} (i.e. if \texttt{C3423567} is the most prolific
shopper of frozen goods, find other customers who are the most
prolific shoppers of fresh produce). The \texttt{analogyUDF()} UDF fetches vectors for the input
variables, and using the 3COSMUL approach, returns the analogy score
between a vector corresponding to the input token and the computed
response vector. Those rows, whose variables (e.g., \texttt{custID})
have analogy score greater than a specified bound (0.5), are selected,
and returned in descending order of the score. Since analogy
operation is implemented using \emph{untyped} vectors, the
\texttt{analogyUDF()} UDF can be used to capture relationships between
variables of different types, e.g., images and text.

\begin{figure}
\hrule
{\small \tt{
\begin{flushleft}
\ix \\
SELECT X.custID, analogyUDF(`Frozen Goods', `custF',`Fresh Produce',X.custID) AS similarity\\
FROM sales X\\
WHERE analogyUDF(`Frozen Goods', `C3423567',`Fresh Produce',X.custID) >
0.5\\
ORDER BY similarity DESC\\
\end{flushleft}
}
\hrule
}
\caption{Example of an analogy query}
\label{fig:analogy-sql}
\end{figure}

\item Semantic Clustering: Given a set of input entities, $\{X,Y,Z,..\}$,
  the semantic  clustering process identifies a set of entities, $\{W,..\}$, that share the
  most dominant trait with the input data. The semantic clustering
 operation has a wide set of applications, including customer
  segmentation, recommendation, etc. Figure~\ref{fig:semcluster-sql}
  presents a CI query which uses a semantic clustering UDF,
  \texttt{semclusterUDF()}, to identify customers that have the most
  common attributes with the input set of customers, e.g.,
  \texttt{custF}, \texttt{custM}, and \texttt{custR}. For solving
a semantic clustering query of the form, $(X,Y,Z::?)$, one needs
to find a set of tokens $S_w=\{W_1,W_2,..,W_i\}$ whose meaning vectors $V_{w_{i}}$ are most similar to the
\emph{centroid} of vectors $V_X,V_Y,$ and $V_Z$ (the centroid vectors
captures the dominant features of the input entities).

\begin{figure}
\hrule
{\small \tt{
\begin{flushleft}
\ix \\
SELECT X.custID, semclusterUDF(`custX', `custY', `custZ', X.custID) AS similarity\\
FROM sales X\\
WHERE semclusterUDF(`custX',`custY', `custZ', X.custID) >
0.5\\
ORDER BY similarity DESC\\
\end{flushleft}
}
\hrule
}
\caption{Example of a semantic clustering query}
\label{fig:semcluster-sql}
\end{figure}

\item Analogy Sequences and Clustered Analogies: These two types of inductive reasoning queries,
analogy sequences and clustered analogies, can be implemented by
combining strategies for semantic clustering and analogies. The
analogy sequence query takes as input a sequence of analogy pairs,
with the \emph{same} source entity and aims to identify a set of target
entities that exhibit the same relationships with the source entity as the
set of input target entities. To answer this query, one needs to
first compute the centroid vector of the input target vectors and then
use it to answer the following analogy problem:
$source:input\_centroid::source:?$ using the 3COSMUL approach to
return a set of target entities.

Unlike analogy sequences, the
clustered analogy operation takes as input a set of analogy pairs,
each with different $(source_i, target_i)$ entity pairs and aims to predict
a set of $(source_o, target_o)$ pairs that shares the following
relationships with the input sequence: the result set of source entities,
$source_o$, share the dominant trait with the input set of source
entities, $source_i$, and each resultant target entity, $target_o$,  is related to
the corresponding source entity via the analogy
relationship. Therefore, to solve clustered analogy queries, we
first perform semantic clustering to compute the result source
entities, $source_o$, and then for each result source entity, we
compute a set of target entities using the analogy sequence
approach. Unlike the analogy sequences query, the result of the
clustered analogy query is a set of sets: a set of source entities, each associated
with a set of target entities.

\item Odd-man-out\footnote{Odd-man-out is often used in practice to
    quantify human intelligence~\cite{diascro:odd-man-out}.}: As the name suggests, given a set of items, the
  odd-man-out query identifies an item that is semantically different
  from the remaining items~\cite{diascro:odd-man-out}. The odd-man-out
  query can be viewed as a complementary query to semantic
  clustering. For example, given a set of animals, \{Hippopotamus, Giraffe, Elephant, and
  Lion\}, one answer can be Lion, as it is only carnivorous animal
  in the collection. However, if the word-embedding model of these
  animals captures their locations, Elephant may be the answer if it is
  not present in that location. Thus, the odd-man-out execution
  requires context-specific semantic clustering over the meaning
  vectors. Specifically, the clustering aims to partition the data
  into two clusters, one with only one member, and the other
  containing the remaining data. One obvious application of the
  odd-man-out CI query would anomaly detection, e.g., for identifying a
  fraudalant transaction for a customer.

\end{itemize}

\noindent \textbf{(3) Cognitive OLAP Queries:} Figure~\ref{fig:olap}
presents a simple example of using semantic similarities in the
context of a traditional SQL aggregation query. This CI query aims to
extract the maximum sale amount for each product category in the
\texttt{sales} table for each merchant that is similar to a specified
merchant, \texttt{Merchant\_Y}. The result is collated using the values
of the product category. As illustrated earlier, the UDF
\texttt{similarityUDF} can also be used for identifying customers that
are different than the specified merchant. The UDF can use either an
externally trained or locally trained model. This query can be easily
adapted to support other SQL aggregation functions such as
\texttt{MAX()}, \texttt{MIN()}, and \texttt{AVG()}. This query can be
further extended to support \texttt{ROLLUP} operations over the
aggregated values~\cite{Ho:OLAP-range}.

\begin{figure}
\hrule
{\small \tt{
\begin{flushleft}
\ix \\
SELECT X.Category, MAX(X.Amount)\\
FROM sales X\\
WHERE similarityUDF(`Merchant\_Y',X.Merchant) >
0.5\\
GROUP BY X.Category\\
\end{flushleft}
}
\hrule
}
\caption{Example of a cognitive OLAP (aggregation) query}
\label{fig:olap}
\end{figure}

We are also exploring integration of cognitive capabilities into
additional SQL operators, e.g., \texttt{IN} and \texttt{BETWEEN}. For
example, one or both of the value ranges for the \texttt{BETWEEN}
operator can be computed using a similarity CI query. For an
\texttt{IN} query, the associated set of choices can be generated by a
similarity or inductive reasoning queries. Another intriguing
extension involves using contextual similarities to choose
\emph{members} of the schema dimension hierarchy for aggregation
operations like \texttt{ROLLUP} or \texttt{CUBE}. For example, instead
of aggregating over all quarters for all years, one can use only those
quarters that are semantically similar to a specified quarter.

\noindent \textbf{(4) Cognitive Extensions to the Relational Data
  Model:} There are powerful extensions to SQL that are enabled by word
vectors. For this we need the ability to refer to constituent tokens
(extracted during textification) in columns of rows, in whole rows and
in whole relations. The extension is via a declaration, in the \texttt{FROM}
clause, of the form \texttt{Token e1} that states that variable $e1$
refers to a token. To \emph{locate} a token we use, in the \texttt{WHERE} clause,
predicates of the form \texttt{contains(E, e1)} where \texttt{E} can be a
column in a row (e.g., \texttt{EMP.Address}), a whole row (e.g.,
\texttt{EMP.*}) or a whole relation (e.g., \texttt{EMP}). With this extension
we can easily express queries such as asking for an employee whose
Address contains a token which is very close to a token in a row in
the \texttt{DEPT} relation (Figure~\ref{fig:sql-entity1}). 
Furthermore, we can also extend SQL with relational variables, say of
the form \texttt{\$R} and column variables, say \texttt{X}, whose names are not
specified at query writing time; they are bound at runtime. We can
then use these variables in queries, in conjunction with \texttt{Token} variables.
This enables database querying without explicit schema knowledge which
is useful for exploring a database. 
Interestingly, the notation \texttt{\$R.X} is basically syntactic sugar.
A software translation tool can substitute for \texttt{\$R.X} an actual
table name and an actual column. Then, perform the query for each such
substitution and return the \emph{union} of the
results~\cite{bordawekar:corr-abs-1603-07185}.

\begin{figure}[htbp]
\hrule
{\small \tt{
\ix \\
SELECT EMP.Name, EMP.Salary, DEPT.Name \\
FROM EMP, DEPT, Token e1, e2\\
WHERE contains(EMP.Address, e1) AND\\
contains(DEPT.*, e2) AND\\
cosineDistance(e1, e2) > 0.75\\
}
\hrule
}
\caption{Example of an SQL query with entities}
\label{fig:sql-entity1}
\end{figure}

Lastly, one may wonder how numeric bounds on UDFs (e.g.,
\texttt{cosineDistance(e1, e2) > 0.75} in
Figure~\ref{fig:sql-entity1}) are determined. The short answer is that
these bounds are application dependent, much like hyperparameters in
machine learning.  One learns these by exploring the underlying
database and running experiments. In the future, one can envision a tool
that can guide users to select an appropriate numeric bound for a particular CI query.

\section{Cognitive Queries in Practice}
\label{sec:practice}

\begin{figure}[htbp]
  \centering
    \includegraphics[width=0.5\textwidth]{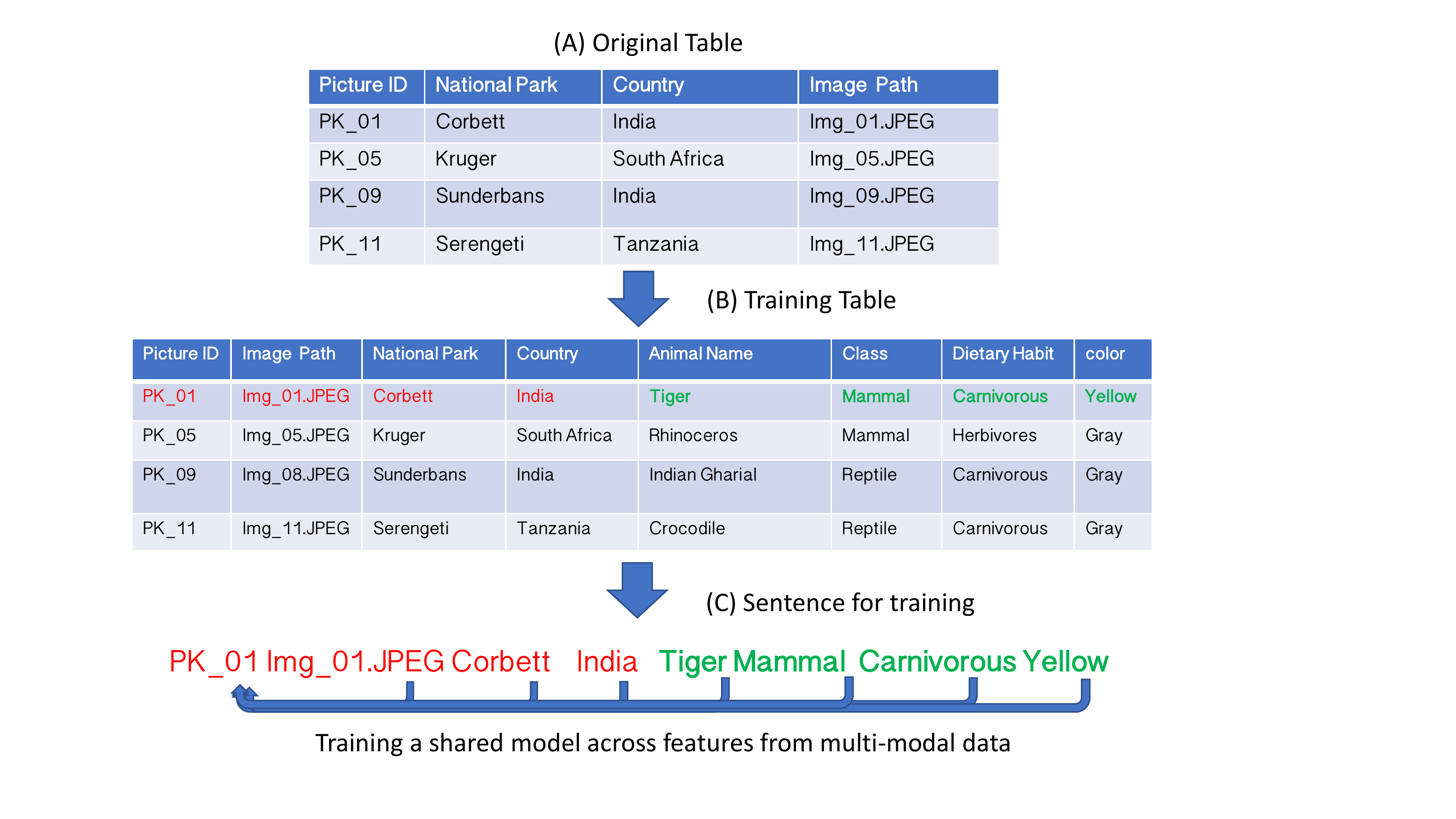}
        \caption{Steps in training a multi-modal database with text
          and image fields} 
        \label{fig:imgtrain}
\end{figure}

We now illustrate some unique capabilities of our cognitive database system by
discussing a scenario in which CI queries are used
to gain novel insights from a multi-modal relational database.\footnote{Due to space limitations, we focus only on certain
  types of queries.} In this scenario, we consider a database of
national parks across multiple countries, with links to images of
animals in the associated national parks (Figure~\ref{fig:imgtrain}). We use images from the open
source Image database, ImageNet~\cite{imagenet_cvpr09}, to populate our
database. We use this database to present results from inductive
reasoning CI queries using our Spark 2.2.0 based prototype on an Intel
Xeon E5-2680 system. Our
prototype is implemented in Scala and supports queries written in
either Scala or Python using Spark SQL, Spark Dataframes, or Python
pandas SQLite interfaces. Although
the database under evaluation is fairly simple, its architecture is 
similar to many other real life databases, e.g., a multi-modal patient database with
text fields describing patient characteristics and image fields
referring to associated images (e.g., radiology or FMRI images), or an
insurance claims database with text fields containing the claim
information and image fields storing supporting pictures (e.g., car
collision photos).

\begin{figure}[htbp]
  \centering
    \includegraphics[width=0.5\textwidth]{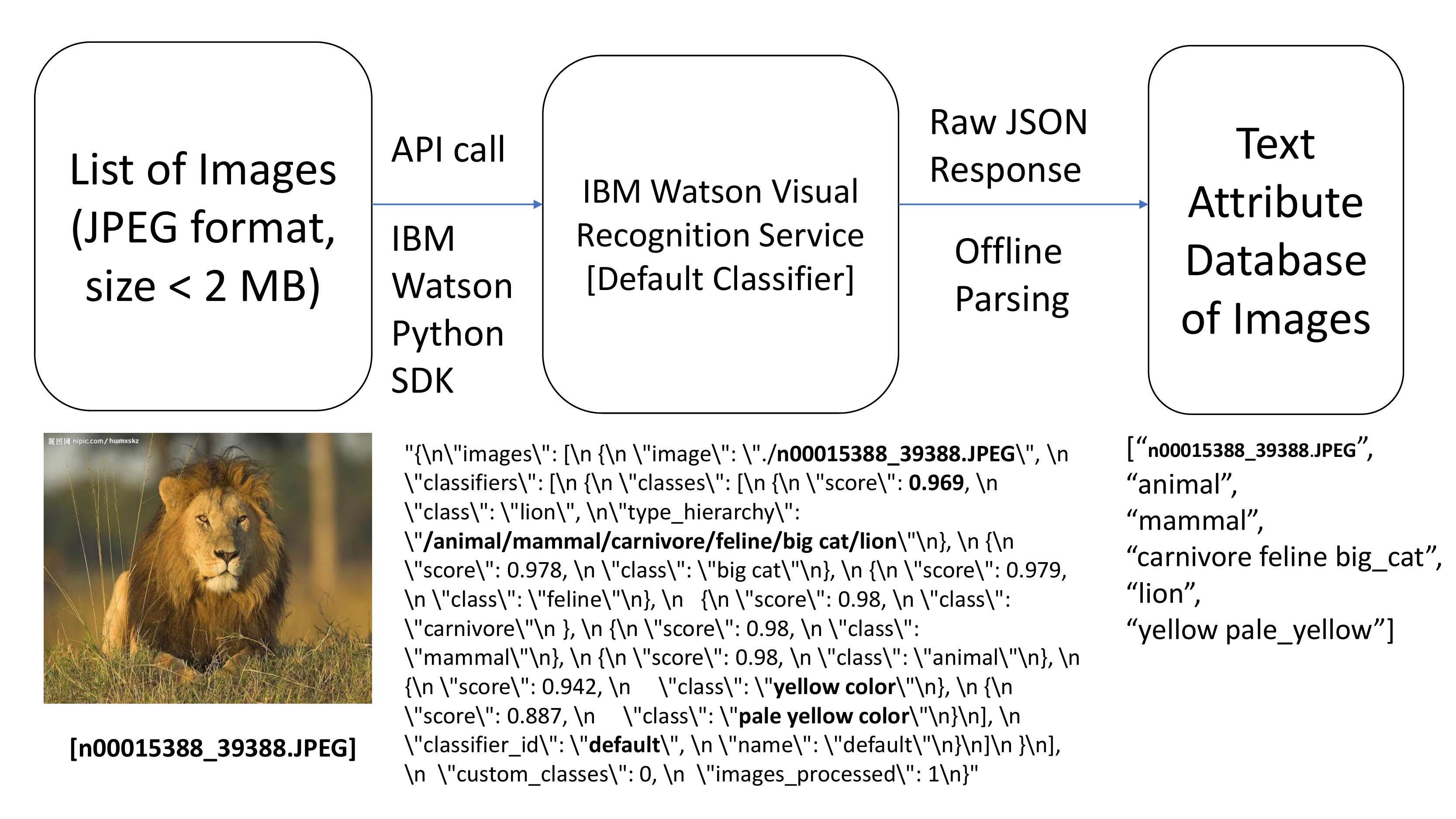}
        \caption{Illustration of workflow to generate a \textit{Text
            Attribute Database} of images using the IBM Watson
            Visual Recognition Service (VRS) using a sample image from
          Imagenet dataset} 
        \label{fig:imgworkflow}
\end{figure}

Figure~\ref{fig:imgtrain}(A) presents the original relational table as
created by an user. It contains only text fields that list paths to
the images and provide additional information on every image. This
table, other than the image path, does not provide any details on the
referred images. To create a shared word embedding model from the text
and images, we employ the automatic tag generator approach outlined in
Section ~\ref{sec:model-prep}. Figure~\ref{fig:imgworkflow} presents
the workflow of extracting image features from the referred
images. Each image in the database (e.g., a Lion) is first uploaded to
the IBM Watson Visual Recognition System (VRS) ~\cite{VRS} for classification and text
description. The Watson system's JSON response is then parsed, and a
set of text attributes for the input image is extracted. These
attributes form the features of the images and are added to the
original table to create a training version of the table (Figure~\ref{fig:imgtrain}(B)). This
training table can be either hidden or exposed to the user. The training
table is then converted to the textual representation
(Figure~\ref{fig:imgtrain}(C)) to build the text corpus for training
the word embedding model. Each sentence in the text corpus includes
both original non-image (e.g., \texttt{Corbett}) and extracted image
features (e.g., \texttt{Tiger}, or \texttt{Carnivorous}). In the
resulting multi-modal word embedding model, the non-image features will contribute to the meaning
of image features and vice versa, and all meaning vectors will be
uniformly represented using vectors of dimension $d=200$.

Once the meaning vectors are computed they can be used to evaluate
semantic relationships between values in the \emph{original} relational table. For example, if one were to
compare National Parks, \texttt{Serengati} and \texttt{Sunderbans}
would be the most similar as they both share multiple image
features. It should be noted that one can not get this insight by using
standard SQL queries as the original table does not have any image
information. Further, as many values in the training database are
syntactically different (e.g., \texttt{Crocodile} and \texttt{Indian
  Gharial}), existing SQL systems will fail to extract any semantic
similarities. 

%\begin{figure}
%  \centering
%    \includegraphics[width=0.5\textwidth]{semantic1.pdf}
%        \caption{Semantic Clustering of Images using image object vectors} 
%        \label{fig:semantic1}
%\end{figure}

\begin{figure}[htbp]
  \centering
    \includegraphics[width=0.5\textwidth]{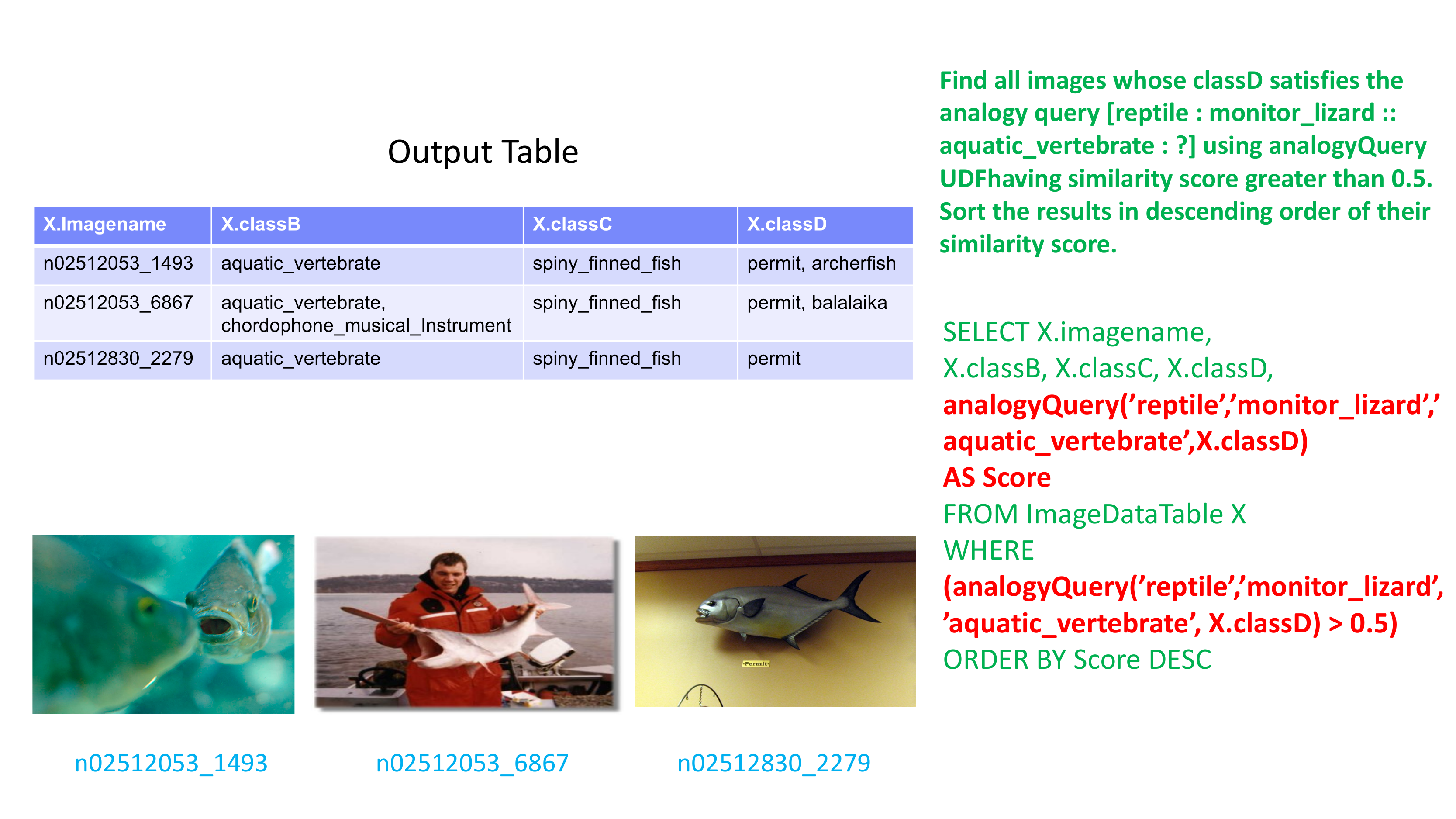}
        \caption{Analogy queries over images} 
        \label{fig:analogy}
\end{figure}

\begin{figure}[htbp]
  \centering
    \includegraphics[width=0.5\textwidth]{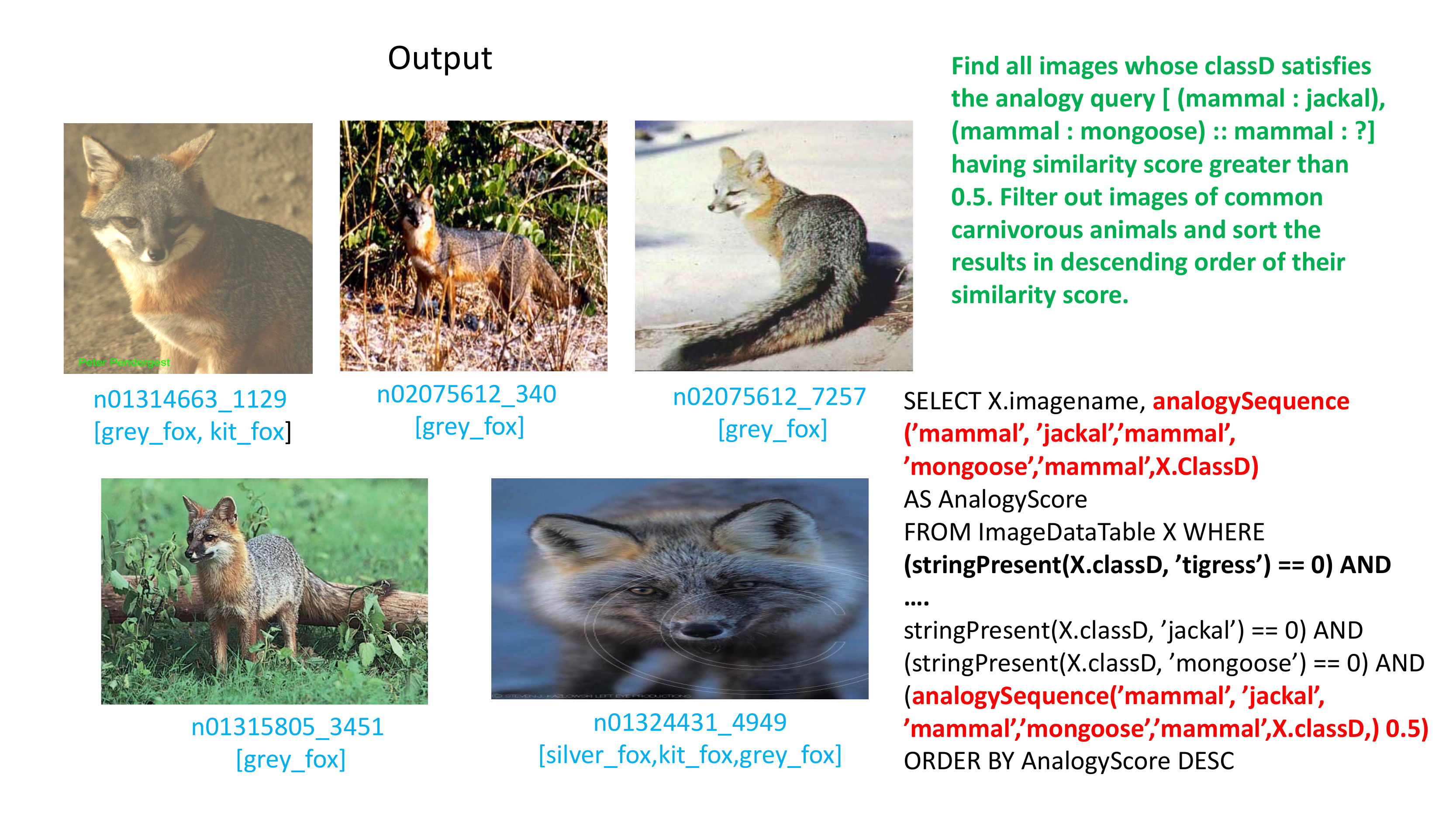}
        \caption{Analogy sequence queries over images} 
        \label{fig:seq}
\end{figure}

The first two examples of SQL CI queries over multi-modal data assume
that the users have access to the training database
(Figure~\ref{fig:imgtrain}(B)). Figure~\ref{fig:analogy} illustrates
an analogy query over images, while Figure~\ref{fig:seq} illustrates
an analogy sequence query. In both cases, the CI queries are formulated
using UDFs, \texttt{analogyQuery()} and \texttt{analogySequence()},
that take values from the training database as input. In case of the
analogy query, the goal is to find all images whose \texttt{classD}
feature (i.e. extracted name) has the same relationship to its
\texttt{classC} feature (i.e. class \texttt{aquatic\_vertebrate}) as
the specified relationship, \texttt{reptile::monitor\_lizard}. For each
row in the table, the UDF first fetches meaning vectors for the input
parameters and uses the 3COSMUL approach to find a relational value
whose vector maximizes the analogy similarity score as defined in
Equation~\ref{3cosmul}.  The SQL query returns the corresponding
images, whose similarity score is higher than 0.5 and reports them in a descending order of similarity
score. Figure~\ref{fig:analogy} presents an output fragment of the CI query
and the corresponding images of \texttt{spiny\_finned\_fish}. 

The analogy sequence query (Figure~\ref{fig:seq}) uses a UDF that
operates on a sequence of analogy pairs that share the source entity (e.g.,
\texttt{mammal}), but has different target entities, (e.g.,
\texttt{jackal} and \texttt{mongoose}). The UDF converts the analogy
sequence problem into a traditional analogy problem by first computing the
average vector of the target entities and then using it in the
3COSMULT approach. Figure~\ref{fig:seq}  presents the SQL CI query,
the analogy sequence UDF, and its output (images of grey fox).

\begin{figure}[htbp]
  \centering
    \includegraphics[width=0.5\textwidth]{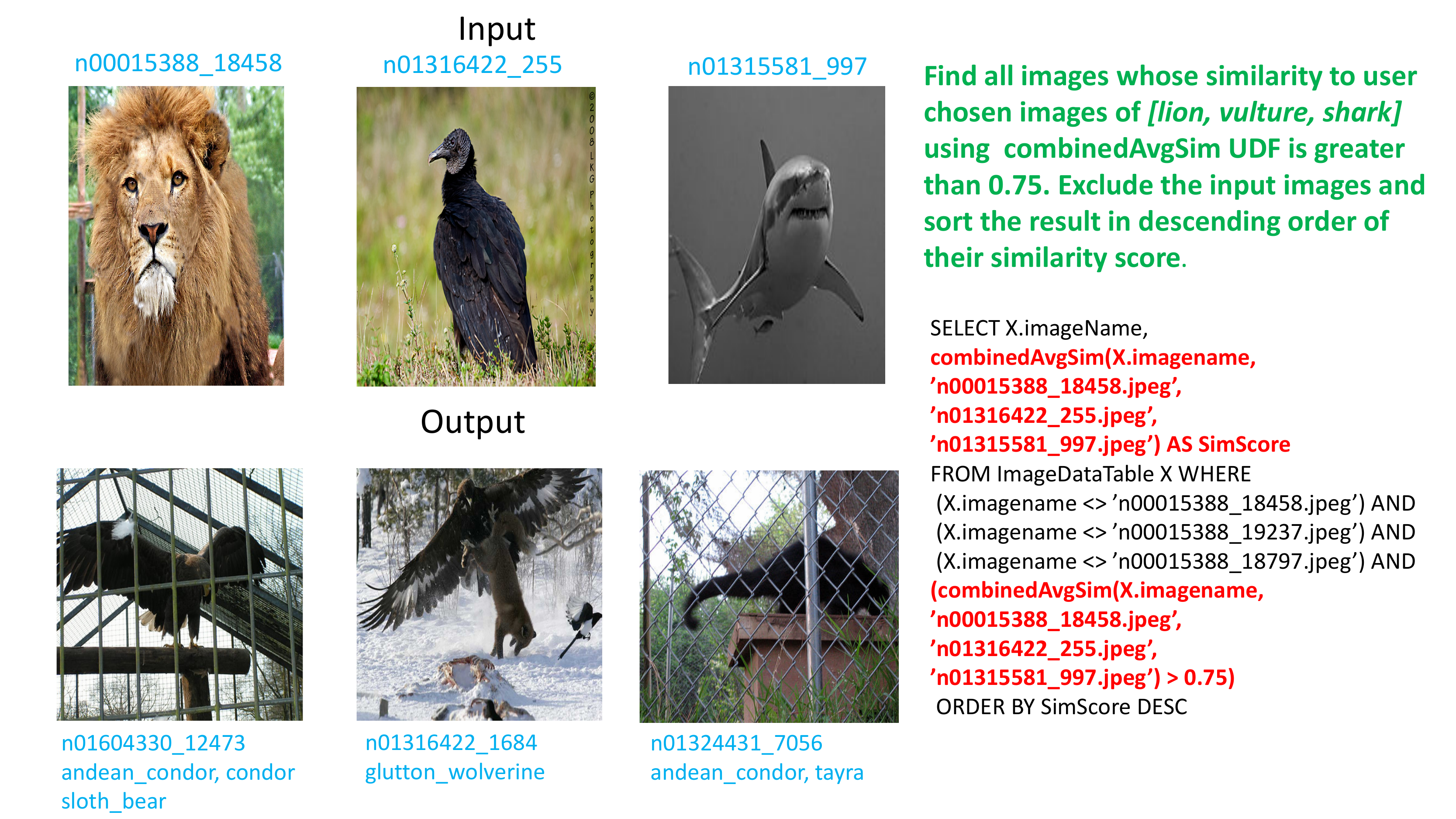}
        \caption{Semantic clustering of images} 
        \label{fig:semantic2}
\end{figure}

The next example (Figure~\ref{fig:semantic2}) illustrates execution of
a semantic clustering query on the \emph{original} multi-modal
database table. The goal of this query is to identify all images that
are similar to every image in the set of user chosen images. Such images share one or
more features with the input set of images. For this query, we select
images of a lion, a vulture, and a shark as the input set and use the
\texttt{combinedAvgSim()} UDF to identify images that are similar to
all these three images. Although the input images display animals
from three different classes, they share one common feature: all three
animals are carnivorous. The UDF computes the average vector from the
three input images and then selects those images whose vectors
are similar to the computed average vector with similarity score
higher than 0.75. Figure~\ref{fig:semantic2} shows the top three image
results: andean condor, glutton wolverine, and tyra. Although these
animals are from different classes, they all are carnivores, a feature
that is shared with the animals from the input set.

\begin{figure}
  \centering
    \includegraphics[width=0.5\textwidth]{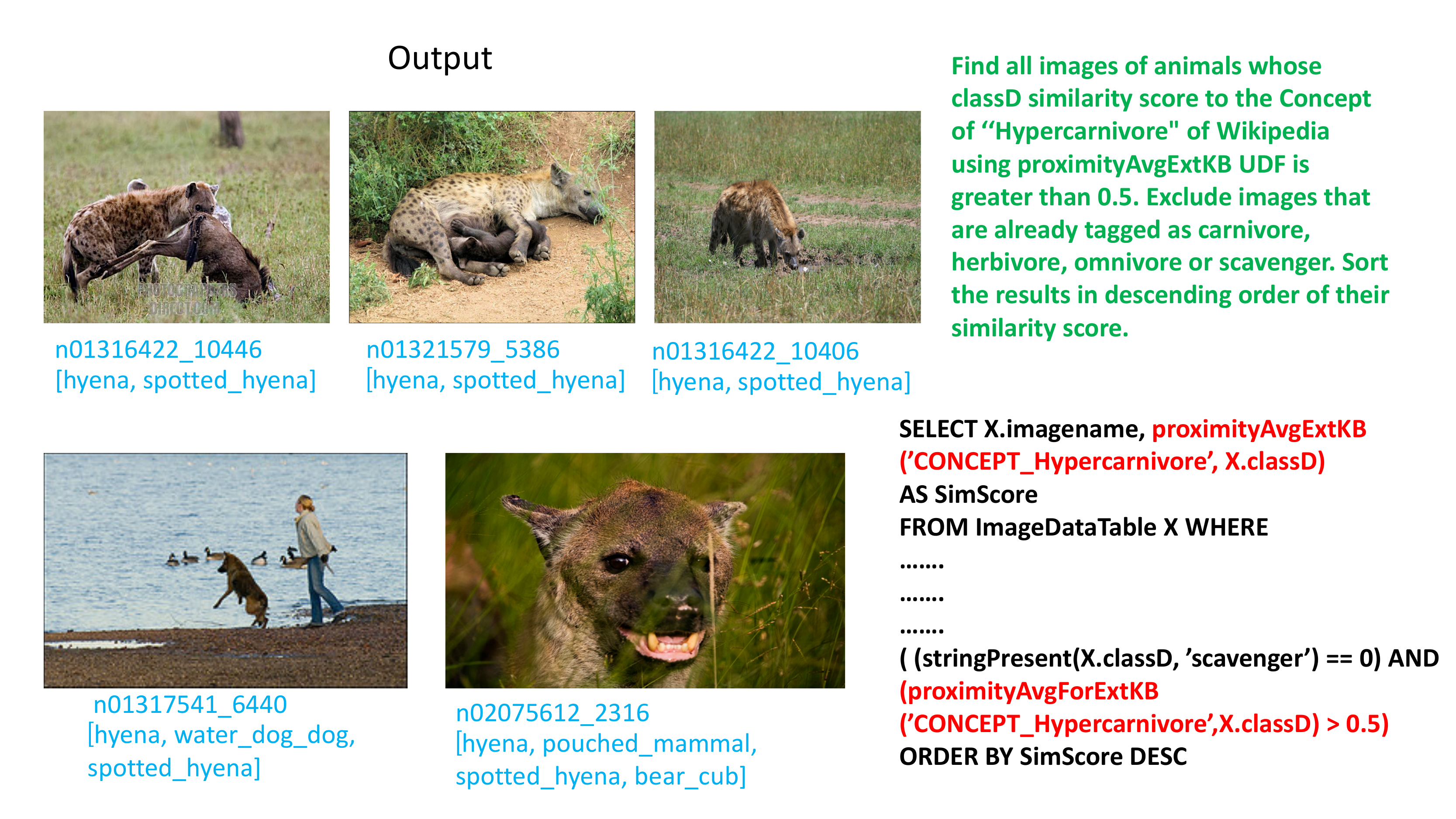}
        \caption{Similarity query over images using a model trained
          from external knowledge base (Wikipedia)} 
        \label{fig:external}
\end{figure}

The final example demonstrates the use of an external semantic model for
querying a multi-modal database. In this scenario, we first train a
word embedding model from an external knowledge base derived from
Wikipedia. Similar to the model trained from the database, the
external model assigns $d$ dimensional meaning vectors to
unique tokens (for the external model, we use $d=200$). From the
wikipedia model, we select a token associated with a concept
\texttt{Hypercarnivore}, which refers to a class of animals whose diet
has more than 70\% meat. Examples of hypercarnivores include lions,
sharks, polar bears, crocodiles, hyenas, etc. Therefore, in our model,
the Hypercarnivore meaning vector is related to meaning vectors of
tokens \texttt{shark}, \texttt{crocodiles}, etc. For this query, we
employ this externally trained model to extract images that are
similar to the concept \texttt{Hypercarnivore}. The UDF
\texttt{proximityAvgForExtKB()} uses the external model, finds
images from the database whose classD features (i.e., names) are
related to \newline \texttt{Hypercarnivore}, and returns those images whose
similarity score is higher than 0.5. Figure~\ref{fig:external} shows
the CI query and its result: pictures of hyenas, who are members of
the hypercarnivore class\footnote{one of the images is a picture of a
  big dog which has been mislabelled as an hyena by Watson VRS.}. This
example also demonstrates the unique capability of cognitive
databases that allows querying a database using a token not present in
the database. In our case, both the original and training databases do
not contain the token \texttt{Hypercarnivore}. 

Although we used hypothetical scenarios to demonstrate our ideas, CI
queries are applicable to a broad class of domains. These include
finance, insurance, retail, customer care, log analytics, healthcare, genomics, semantic
search over documents (patent, legal, or  financial), healthcare
informatics, and human resource management.

These examples demonstrate several unique capabilities of 
cognitive database systems, namely: (1) ability to build joint cross-modal
semantic model from multi-modal data, (2) \emph{transperent} integration
of novel cognitive queries into existing SQL query infrastructure, (3)
using untyped vector representations to support contextual semantic
(i.e. not value based) queries across multiple SQL data types, and
(4) ability to import an externally trained semantic model and apply it
to enable querying a database using a token not present in the
database. To the best of our knowledge, none of the current
industrial, academic, or open source database systems support these capabilities.

\subsection{Optimizing Cognitive Intelligence Queries}

Cognitive Intelligence queries are standard SQL analytics
queries which invoke UDFs to enable contextual semantic operations between
relational variables.  Irrespective of the kind of CI Query, core
UDF computation involves computing pair-wise similarities between
vectors, which can be then used to identify nearest or
furthest neighbors of a vector. In the worst case, a CI query can
invoke a UDF for every row combination being evaluated and the UDF, in turn, can operate on
a large number of vectors. Since in practice, the number of row combinations can very high, it is critical
to optimize the performance of distance computations in the
nearest-neighbor calculations for CI queries.

In a $d$ dimensional vector space, pair-wise distance between vectors
$v_1$ and $v_2$ is calculated by computing the cosine distance,
$cos(v_1,v_2)=\frac{v_1\cdot v_2}{\lVert {v_1} \rVert \lVert
  {v_2} \rVert}$. As we use normalized forms
of vectors (i.e. $\lVert v \rVert=1$), the pair-wise distance calculation
gets simplified to a vector dotproduct, $v_1\cdot v_2$. In general,
there are four basic ways of optimizing distance computations:

\begin{itemize}

\item Increasing computation granularity:  In cases
where one needs to compute the distance of a vector from many vectors,
many individual dotproduct operations can be converted into a single
matrix-vector multiplication operation. This can be generalized to a
matrix-matrix multiplication operation to enable distance computations
between two sets of vectors.

\item Reducing redundant computations: For a given
  vector, we can first identify a candidate set of vectors that are spatially closer to it in the $d$
  dimensional vector space using either locality sensitive hashing
  (LSH)~\cite{srp-charikar}, or clustering via the Spherical K-Means
  algorithm~\cite{dhillon-kmeans}. We can then invoke distance calculations on the candidate set to compute precise distances.
  In the LSH approach, the $d$ dimension vector locations are mapped
  to bit-vector \emph{signatures} of length
  $d$ via projecting them on random planes. For a given vector, spatially closer vectors can be identified
  by choosing those with small hamming distance (1 or 2) between their
  corresponding signatures. In the K-Means approach, one can use
  the centroid information to identify a candidate set of spatially close vectors. Both approaches can also be
  accelerated using either SIMD functions or GPUs~\cite{nn-gpu}.

\item Using relational query optimization approaches: One can also
  reduce redundant computations by using relational view that
  pre-selects rows from a table based on certain criteria. In
  addition, once the candidate set of SQL variables is known, the SQL
  query engine can pre-compute the pair-wise distances and use the
  cached results during execution of UDFs later. 

\item Using hardware acceleration and parallelization: The core nearest-neighbor distance computations, namely,
  dotproduct, matrix-vector, and matrix-matrix computations can be
  accelerated via hardware accelerators such as on-chip SIMD or using GPUs~\cite{bordawekar:book}. Most numerical libraries such as
  MKL, ESSL, or OpenBLAS provide hardware accelerated matrix
  computation kernels. Further, the nearest neighbor computations can
  be also parallelized either using CPU-based multithreading (e.g.,
  using pthreads) or distributing it over a cluster of machines using
  a distributed infrastructure such as the Apache Spark.

\end{itemize}

Optimizations of the cognitive intelligence queries is an open
problem and is the current focus of our activities.

\section{Related Work}
\label{sec:related}

\noindent \textbf{Language Embedding:}
Over the past few years, a number of methods have been introduced for
obtaining a vector representation of words in a language 
\cite{bengio:jmlr03}, called \textit{language embedding}. 
The methods range from \emph{brute force} learning by various types of
neural networks (NNs) ~\cite{bengio:jmlr03}, to
log-linear classifiers~\cite{mikolov:corr-abs-1301-3781} and to
various matrix formulations, such as matrix  factorization techniques
\cite{levy:nips14}. Lately, Word2Vec~\cite{w2v, mikolov:nips13,
  mikolov:corr-abs-1309-4168, levy:conll14} has
gained prominence as the vectors it produces appear to capture
syntactic as well semantic properties of words.  The exact mechanism
employed by Word2Vec and suggestions for alternatives are 
the subject of much research~\cite{mnih-nips13,
  goldberg:corr-abs-1402-3722,
  levy:conll14}. Although  Word2Vec has gained much prominence
it is one of many possible methods for generating word representing vectors. For example,
GloVe~\cite{pennington:glove14} also builds word embeddings by a
function optimization approach over the word co-occurrence matrix. Vectors may be associated with larger bodies of text such as
 paragraphs and even documents. Applications to the paragraph and
 document embedding appear in \cite{le:corr14, dai:corr15}. Recent
 work has also been exploring applying word embeddings to capture
 image semantics~\cite{socher:zero-shot}.

\noindent \textbf{Applications of Word Embedding:} The word embedding
model is being used for a wide variety of applications beyond NLP. Wu et
al~\cite{wu:starspace} provide a general neural framework for
building vector embaddings of entities of different types into a
vectorial embedding space. The common representation can be then used
for different tasks such as text classification, link prediction,
document recommendation, etc. The YouTube recommendation system uses
embedding approaches to capture user
behavior~\cite{covington:youtube}. Similar embedding-based approaches 
have been used for recommending news articles~\cite{okura:news}, for
discovering topics~\cite{xun:topic}, or for personalized
fashion shopping~\cite{moody:w2v,arora:myntra}. Hope et. al. have proposed using word 
embedding for supporting analogy queries over knowledge bases such as
the US Patent database~\cite{hope:analogy}.
DeepWalk~\cite{perozzi:deepwalk} and Node2Vec~\cite{Grover:Node2Vec} have proposed using word embedding
approaches for learning neighborhood features of nodes in a network
graph. Word embedding approaches are also being used for a variety of
semantic web applications, e.g., embedding RDF
triples~\cite{ristoski:rdf, cochez:RDF} and encoding geo-spatial proximity~\cite{kejriwal:neural}. Using latent feature vectors for data integration in knowledge
databases has been explored in \cite{vilnis:corr-1412-6623, verga:corr-1511-06396}.

%%% Related work in the database world

\noindent \textbf{Relational Databases:}
In the context of SQL, text capabilities, e.g., the use of synonyms,
have been in practice for a while \cite{cutlip:db2}.
In the literature, techniques for detecting similarity between records
and fields have also been explored. Semantic similarity between
database records is explored in~\cite{kashyap:vldbj96}.
Phrase-based ranking by applying an IR approach to relational
data appears in~\cite{liu:sigmod06}.
Indexing and searching relational data by modeling tuples as virtual
documents appear in~\cite{luo:sigmod07}.
Effective keyword-based selection of relational
databases is explored in~\cite{yu:sigmod07}.
A system for detecting XML similarity, in content and structure, using
a relational database is described in~\cite{viyanon:cikm09}.
Related work on similarity Join appears in~\cite{chaudhuri:ssjoin}.
Semantic Queries are described in~\cite{pan:dldb}.
Most recently, Shin et al. have described DeepDive~\cite{Shin:vldb} that
uses machine learning techniques, e.g., Markov Logic based rules, to
convert input unstructured documents into a structured knowledge base.

The proposed cognitive database system can be distinguished by the
following unique features: (1) Encoding relational data using word
embedding techniques, (2) Using semantic vectors to enable a new
class of SQL analytics queries (CI queries), (3) Ability to make
\emph{contextual semantic} matching, unlike the traditional
\emph{value (syntactical)} matching supported by current SQL queries, (4) Capturing
relationalships across multiple data types, including images, and (5)
Ability of using external knowledge bases.  Further, semantic vectors are primarily based on the database \textit{itself}
(with external text or vectors as an option). This means that we
assume no reliance on dictionaries, thesauri, word nets and the
like. Once these vectors are generated they may be used in vastly
enriching the querying expressiveness of virtually any query
language. These capabilities go far beyond analytical capabilities present in current relational systems. All well-known commercial and
open source (e.g., Apache Spark~\cite{Spark}, MADlib~\cite{hellerstein:madlib}) database systems have built-in
analytics capabilities, e.g., Spark MLLib. Apache Spark can also create a deep-learning pipeline
in which it can invoke an external deep-learning infrastructure e.g.,
TensorFlow~\cite{tensorflow} to train a model, and then load the trained model to perform
inferencing operations~\cite{spark-dl}. However, such systems view databases as 
repositories for storing input features and results for the analytics or deep-learning frameworks. On the other hand,
cognitive databases use the word embedding model to extract features
from the database entities and use them to enhance its querying capabilities. Systems based on
statistical relational learning models combine probabilistic graphical models and first-order logic to
encode uncertain first-order logic rules based on \emph{known}
information \cite{wang:graphical}. In contrast, a cognitive database
learns information about the relational data which is not known
apriori.

\section{Conclusions and Success Criteria}
\label{sec:concl}

In this paper we presented Cognitive Database, an innovative
relational database system that uses the power of word embedding
models to enable novel AI capabilities in database systems. The  word embedding approach uses
unsupervised learning  to generate meaning vectors using
database-derived text. These vectors capture syntactic as well as 
semantic characteristics of every database token. We use these vectors to enhance database querying capabilities. In
essence, these vectors provide another way to look at the database,
almost orthogonal to the structured relational regime, as
vectors enable a dual view of the data: relational and meaningful
text. We thereby introduce and explore a new class of queries called
\emph{cognitive intelligence (CI)} queries that extract information
from the database based, in part, on the relationships encoded by
these vectors.

We are implementing a prototype system on top of Apache Spark \cite{Spark} to
exhibit the power of CI queries. Our current infrastructure enables
complex SQL-based semantic queries over multi-modal databases (e.g.,
inductive reasoning queries over a text and image database). We are
now working on accelerating model training and nearest neighbor computations using a
variety of approaches (e.g., using GPUs), and developing new techniques for incremental
vector training. We believe CI queries are applicable to a broad class of
application domains including healthcare, bio-informatics, document
searching, retail analysis, and data integration.

\subsection{Success Criteria} We believe Cognitive Databases are truly a new technology for incorporating 
AI capabilities into relational databases. As such it holds a great
promise for innovative applications with a very different view of data
as compared to today's database systems. Since it is based on a new
concept, there are no easy comparisons. For example, there are no
relevant benchmarks except for ones used in NLP for testing language features
such as analogies \cite{mikolov:nips13}. So, success of  the cognitive databases
will be mainly evaluated based on new applications within known
domains (such as Retail), new domains of applications (such
as medical drugs selection) that are currently in the sphere of
AI-based systems,  and the adaptation of CI capabilities as a standard
feature by leading vendor as well as open source database
systems. Finally, we hope this work spurs new research
initiatives in this exciting emerging area in database management
systems.

\newpage

\bibliographystyle{abbrv}
\bibliography{related}

\begin{thebibliography}{10}

\bibitem{Spark}
{Apache Foundation}.
\newblock Apache spark: A fast and general engine for large-scale data
  processing, 2017.
\newblock Release 2.2.

\bibitem{arora:myntra}
S.~Arora and D.~Warrier.
\newblock Decoding fashion contexts using word embeddings.
\newblock In {\em KDD Workshop on Machine learning meets fashion}, 2016.

\bibitem{bahdanau:attention}
D.~Bahdanau, K.~Cho, and Y.~Bengio.
\newblock Neural machine translation by jointly learning to align and
  translate.
\newblock In {\em Proceedings of the ICLR 2015}, 2015.

\bibitem{baroni:compare}
M.~Baroni, G.~Dinu, and G.~Kruszewski.
\newblock Don’t count, predict! a systematic comparison of context-counting
  vs. context-predicting semantic vectors.
\newblock In {\em Proceedings of The 2014 Conference of the Association for
  Computational Linguistics}, 2014.

\bibitem{bengio:jmlr03}
Y.~Bengio, R.~Ducharme, P.~Vincent, and C.~Janvin.
\newblock A neural probabilistic language model.
\newblock {\em Journal of Machine Learning Research}, 3:1137--1155, 2003.

\bibitem{bordawekar:book}
R.~Bordawekar, B.~Blainey, and R.~Puri.
\newblock {\em Analyzing Analytics}.
\newblock Morgan \& Claypool Publishers, November 2015.
\newblock Synthesis Lectures on Computer Architecture.

\bibitem{nn-gpu}
R.~Bordawekar and P.~D'Souza.
\newblock {Optimizing Out-of-core Nearest Neighbor Problems on Multi-GPU
  Systems using NVLink}.
\newblock Nvidia Global Technical Conference (GTC), May 2017.

\bibitem{bordawekar:corr-abs-1603-07185}
R.~Bordawekar and O.~Shmueli.
\newblock Enabling cognitive intelligence queries in relational databases using
  low-dimensional word embeddings.
\newblock {\em CoRR}, abs/1603.07185, March 2016.

\bibitem{srp-charikar}
M.~S. Charikar.
\newblock Similarity estimation techniques from rounding algorithms.
\newblock In {\em Proceedings of the Thiry-fourth Annual ACM Symposium on
  Theory of Computing}, pages 380--388, 2002.

\bibitem{chaudhuri:ssjoin}
S.~Chaudhuri, V.~Ganti, and R.~Kaushik.
\newblock A primitive operator for similarity joins in data cleaning.
\newblock In {\em Proceedings of the 22nd International Conference on Data
  Engineering}, Washington, DC, USA, 2006. IEEE Computer Society.

\bibitem{chu:relax}
W.~W. Chu, H.~Yang, and G.~Chow.
\newblock A cooperative database system (cobase) for query relaxation.
\newblock In {\em Proceedings of ARPI 1996}, 1996.

\bibitem{cochez:RDF}
M.~Cochez, P.~Ristoski, S.~P. Ponzetto, and H.~Paulheim.
\newblock Global rdf vector space embeddings.
\newblock In {\em Proceedings of the 16th International Semantic Web
  Conference}, 2017.

\bibitem{covington:youtube}
P.~Covington, J.~Adams, and E.~Sargin.
\newblock Deep neural networks for youtube recommendations.
\newblock In {\em Proceedings of the 10th ACM Conference on Recommender
  Systems}, New York, NY, USA, 2016.

\bibitem{cutlip:db2}
R.~Cutlip and J.~Medicke.
\newblock {\em Integrated Solutions with DB2}.
\newblock IBM Press, 2003.

\bibitem{dai:corr15}
A.~M. Dai, C.~Olah, and Q.~V. Le.
\newblock Document embedding with paragraph vectors.
\newblock {\em CoRR}, abs/1507.07998, 2015.

\bibitem{imagenet_cvpr09}
J.~Deng, W.~Dong, R.~Socher, L.-J. Li, K.~Li, and L.~Fei-Fei.
\newblock {ImageNet: A Large-Scale Hierarchical Image Database}.
\newblock In {\em CVPR09}, 2009.

\bibitem{dhillon-kmeans}
I.~S. Dhillon and D.~S. Modha.
\newblock Concept decompositions for large sparse text data using clustering.
\newblock {\em Mach. Learn.}, 42(1-2):143--175, Jan. 2001.

\bibitem{diascro:odd-man-out}
M.~N. Diascro and N.~Brody.
\newblock Odd-man-out and intelligence.
\newblock {\em Intelligence}, 19(1):79--92, July-August 1994.

\bibitem{goldberg:corr-abs-1402-3722}
Y.~Goldberg and O.~Levy.
\newblock word2vec explained: deriving mikolov et al.'s negative-sampling
  word-embedding method.
\newblock {\em CoRR}, abs/1402.3722, 2014.

\bibitem{goodfellow-deep}
I.~Goodfellow, Y.~Bengio, and A.~Courville.
\newblock {\em Deep Learning}.
\newblock The MIT Press, 2016.

\bibitem{tensorflow}
{Google Inc.}
\newblock {TensorFlow: An open-source software library for Machine
  Intelligence}.

\bibitem{gray:olap}
J.~Gray, S.~Chaudhuri, A.~Bosworth, A.~Layman, D.~Reichart, M.~Venkatrao,
  F.~Pellow, and H.~Pirahesh.
\newblock Data cube: A relational aggregation operator generalizing group-by,
  cross-tab, and sub-totals.
\newblock {\em Data Mining and Knowledge Discovery}, 1(1):29--53, 1997.

\bibitem{Grover:Node2Vec}
A.~Grover and J.~Leskovec.
\newblock Node2vec: Scalable feature learning for networks.
\newblock In {\em Proceedings of the 22nd ACM SIGKDD International Conference
  on Knowledge Discovery and Data Mining}, KDD '16, pages 855--864, 2016.

\bibitem{harris:distr}
Z.~S. Harris.
\newblock Distributional structure.
\newblock {\em Word}, 10(2-3):146--162, 1954.

\bibitem{hellerstein:madlib}
J.~M. Hellerstein, C.~Re, F.~Schoppmann, D.~Z. Wang, E.~Fratkin, A.~Gorajek,
  K.~S. Ng, C.~Welton, X.~Feng, K.~Li, and A.~Kumar.
\newblock {The MADlib analytics library: or MAD skills, the SQL}.
\newblock {\em Proc. VLDB Endow.}, 5(12), August 2012.

\bibitem{hinton:distr}
G.~E. Hinton.
\newblock Learning distributed representations of concepts.
\newblock In {\em Proceedings of the Eighth Annual Conference of Cognitive
  Science Society}, pages 1--12, 1986.

\bibitem{Ho:OLAP-range}
C.-T. Ho, R.~Agrawal, N.~Megiddo, and R.~Srikant.
\newblock Range queries in olap data cubes.
\newblock In {\em Proceedings of the 1997 ACM SIGMOD International Conference
  on Management of Data}, pages 73--88, 1997.

\bibitem{spark-dl}
S.~A. Hong, T.~Hunter, and R.~Xin.
\newblock A vision for making deep learning simple, June 2017.
\newblock DataBricks Engineering Blog.

\bibitem{hope:analogy}
T.~Hope, J.~Chan, A.~Kittur, and D.~Shahaf.
\newblock Accelerating innovation through analogy mining.
\newblock In {\em Proceedings of the 23rd ACM SIGKDD International Conference
  on Knowledge Discovery and Data Mining}, KDD '17, pages 235--243, 2017.

\bibitem{VRS}
{IBM Watson}.
\newblock Watson visual recognition service.
\newblock www.ibm.com/watson/services/visual-recognition/.

\bibitem{jia2014caffe}
Y.~Jia, E.~Shelhamer, J.~Donahue, S.~Karayev, J.~Long, R.~Girshick,
  S.~Guadarrama, and T.~Darrell.
\newblock Caffe: Convolutional architecture for fast feature embedding.
\newblock {\em arXiv preprint arXiv:1408.5093}, 2014.

\bibitem{kashyap:vldbj96}
V.~Kashyap and A.~P. Sheth.
\newblock Semantic and schematic similarities between database objects: {A}
  context-based approach.
\newblock {\em {VLDB} J.}, 5(4):276--304, 1996.

\bibitem{kejriwal:neural}
M.~Kejriwal and P.~Szekely.
\newblock Neural embeddings for populated geonames locations.
\newblock In {\em Proceedings of the 16th International Semantic Web
  Conference}, 2017.

\bibitem{le:corr14}
Q.~V. Le and T.~Mikolov.
\newblock Distributed representations of sentences and documents.
\newblock {\em CoRR}, abs/1405.4053, 2014.

\bibitem{levy:conll14}
O.~Levy and Y.~Goldberg.
\newblock Linguistic regularities in sparse and explicit word representations.
\newblock In {\em Proceedings of the Eighteenth Conference on Computational
  Natural Language Learning, CoNLL 2014}, pages 171--180, 2014.

\bibitem{levy:nips14}
O.~Levy and Y.~Goldberg.
\newblock Neural word embedding as implicit matrix factorization.
\newblock In {\em Annual Conference on Neural Information Processing Systems
  2014}, pages 2177--2185, 2014.

\bibitem{Li:nlq}
F.~Li and H.~V. Jagadish.
\newblock Constructing an interactive natural language interface for relational
  databases.
\newblock {\em Proc. VLDB Endow.}, 8(1):73--84, Sept. 2014.

\bibitem{lim:edbt13}
L.~Lim, H.~Wang, and M.~Wang.
\newblock Semantic queries by example.
\newblock In {\em Proceedings of the 16th International Conference on Extending
  Database Technology (EDBT 2013)}, 2013.

\bibitem{linzen2016issues}
T.~Linzen.
\newblock Issues in evaluating semantic spaces using word analogies.
\newblock {\em arXiv preprint arXiv:1606.07736}, 2016.

\bibitem{liu:sigmod06}
F.~Liu, C.~T. Yu, W.~Meng, and A.~Chowdhury.
\newblock Effective keyword search in relational databases.
\newblock In {\em Proceedings of the {ACM} {SIGMOD} International Conference on
  Management of Data, Chicago, Illinois, USA, June 27-29, 2006}, pages
  563--574, 2006.

\bibitem{luo:sigmod07}
Y.~Luo, X.~Lin, W.~Wang, and X.~Zhou.
\newblock Spark: top-k keyword query in relational databases.
\newblock In {\em Proceedings of the {ACM} {SIGMOD} International Conference on
  Management of Data, Beijing, China, June 12-14, 2007}, pages 115--126, 2007.

\bibitem{w2v}
T.~Mikolov.
\newblock word2vec: Tool for computing continuous distributed representations
  of words, 2013.
\newblock github.com/tmikolov/word2vec.

\bibitem{mikolov:corr-abs-1301-3781}
T.~Mikolov, K.~Chen, G.~Corrado, and J.~Dean.
\newblock Efficient estimation of word representations in vector space.
\newblock {\em CoRR}, abs/1301.3781, 2013.

\bibitem{mikolov:corr-abs-1309-4168}
T.~Mikolov, Q.~V. Le, and I.~Sutskever.
\newblock Exploiting similarities among languages for machine translation.
\newblock {\em CoRR}, abs/1309.4168, 2013.

\bibitem{mikolov:nips13}
T.~Mikolov, I.~Sutskever, K.~Chen, G.~S. Corrado, and J.~Dean.
\newblock Distributed representations of words and phrases and their
  compositionality.
\newblock In {\em 27th Annual Conference on Neural Information Processing
  Systems 2013.}, pages 3111--3119, 2013.

\bibitem{mnih-nips13}
A.~Mnih and K.~Kavukcuoglu.
\newblock Learning word embeddings efficiently with noise-contrastive
  estimation.
\newblock In {\em 27th Annual Conference on Neural Information Processing
  Systems 2013.}, pages 2265--2273, 2013.

\bibitem{moody:w2v}
C.~Moody.
\newblock A word is worth a thousand vectors.
\newblock {Stitch Fix Multithreaded Blog}.

\bibitem{okura:news}
S.~Okura, Y.~Tagami, S.~Ono, and A.~Tajima.
\newblock Embedding-based news recommendation for millions of users.
\newblock In {\em Proceedings of the 23rd ACM SIGKDD International Conference
  on Knowledge Discovery and Data Mining}, KDD '17, pages 1933--1942, 2017.

\bibitem{pan:dldb}
Z.~Pan and J.~Heflin.
\newblock {DLDB:} extending relational databases to support semantic web
  queries.
\newblock In {\em {PSSS1} - Practical and Scalable Semantic Systems}, 2003.

\bibitem{Park:sigmod17}
Y.~Park, A.~S. Tajik, M.~Cafarella, and B.~Mozafari.
\newblock Database learning: Toward a database that becomes smarter every time.
\newblock In {\em Proceedings of the 2017 ACM International Conference on
  Management of Data}, pages 587--602, 2017.

\bibitem{pennington:glove14}
J.~Pennington, R.~Socher, and C.~D. Manning.
\newblock {GloVe}: Global vectors for word representation.
\newblock In {\em Proceedings of the 2014 Conference on Empirical Methods in
  Natural Language Processing}, pages 1532--1543, 2014.

\bibitem{perozzi:deepwalk}
B.~Perozzi, R.~Al-Rfou, and S.~Skiena.
\newblock Deepwalk: Online learning of social representations.
\newblock In {\em Proceedings of the 20th ACM SIGKDD International Conference
  on Knowledge Discovery and Data Mining}, KDD '14, pages 701--710, 2014.

\bibitem{ristoski:rdf}
P.~Ristoski and H.~Paulheim.
\newblock Rdf2vec: Rdf graph embedding for data mining.
\newblock In {\em Proceedings of the 15th International Semantic Web
  Conference}, pages 498--514, 2016.

\bibitem{Rumelhart1973}
D.~E. Rumelhart and A.~A. Abrahamson.
\newblock A model for analogical reasoning.
\newblock {\em Cognitive Psychology}, 5(1):1 -- 28, 1973.

\bibitem{salton:vsm}
G.~Salton, A.~Wong, and C.~S. Yang.
\newblock A vector space model for automatic indexing.
\newblock {\em Communications of the ACM}, 18(11):613--620, 1975.

\bibitem{Shin:vldb}
J.~Shin, S.~Wu, F.~Wang, C.~De~Sa, C.~Zhang, and C.~R{\'e}.
\newblock Incremental knowledge base construction using deepdive.
\newblock {\em Proc. VLDB Endow.}, 8(11), July 2015.

\bibitem{socher:zero-shot}
R.~Socher, M.~Ganjoo, H.~Sridhar, O.~Bastani, C.~D. Manning, and A.~Y. Ng.
\newblock Zero-shot learning through cross-modal transfer.
\newblock {\em CoRR}, abs/1301.3666, 2013.

\bibitem{Sternberg1979}
R.~J. Sternberg and M.~K. Gardner.
\newblock Unities in inductive reasoning.
\newblock Technical Report Technical rept. no. 18, 1 Jul-30 Sep 79, Yale
  University, 1979.

\bibitem{VanAken:sigmod17}
D.~Van~Aken, A.~Pavlo, G.~J. Gordon, and B.~Zhang.
\newblock Automatic database management system tuning through large-scale
  machine learning.
\newblock In {\em Proceedings of the 2017 ACM International Conference on
  Management of Data}, pages 1009--1024, 2017.

\bibitem{verga:corr-1511-06396}
P.~Verga, D.~Belanger, E.~Strubell, B.~Roth, and A.~McCallum.
\newblock Multilingual relation extraction using compositional universal
  schema.
\newblock {\em CoRR}, abs/1511.06396, 2015.

\bibitem{vilnis:corr-1412-6623}
L.~Vilnis and A.~McCallum.
\newblock Word representations via gaussian embedding.
\newblock {\em CoRR}, abs/1412.6623, 2014.

\bibitem{viyanon:cikm09}
W.~Viyanon and S.~K. Madria.
\newblock A system for detecting xml similarity in content and structure using
  relational database.
\newblock In {\em Proceedings of the 18th {ACM} Conference on Information and
  Knowledge Management, {CIKM}}, pages 1197--1206, 2009.

\bibitem{wang:graphical}
D.~Z. Wang, Y.~Chen, C.~Grant, and K.~Li.
\newblock Efficient in-database analytics with graphical models.
\newblock {\em IEEE Data Engineering Bulletin}, 2014.

\bibitem{wu:starspace}
L.~Wu, A.~Fisch, S.~Chopra, K.~Adams, A.~Bordes, and J.~Weston.
\newblock Starspace: Embed all the things!
\newblock {\em CoRR}, abs/1709.03856, 2017.

\bibitem{xun:topic}
G.~Xun, Y.~Li, J.~Gao, and A.~Zhang.
\newblock Collaboratively improving topic discovery and word embeddings by
  coordinating global and local contexts.
\newblock In {\em Proceedings of the 23rd ACM SIGKDD International Conference
  on Knowledge Discovery and Data Mining}, KDD '17, pages 535--543, 2017.

\bibitem{yu:sigmod07}
B.~Yu, G.~Li, K.~R. Sollins, and A.~K.~H. Tung.
\newblock Effective keyword-based selection of relational databases.
\newblock In {\em Proceedings of the {ACM} {SIGMOD} International Conference on
  Management of Data, Beijing, China, June 12-14, 2007}, pages 139--150, 2007.

\end{thebibliography}

\balance
\end{document}